\documentclass[letterpaper,twocolumn,10pt]{article}

\newif\ifisdraft
\isdraftfalse

\usepackage{usenix2019_v3}

\usepackage[normalem]{ulem}

\usepackage[T1]{fontenc}
\usepackage[utf8]{inputenc}

\usepackage{amsthm}
\usepackage{scalerel, amssymb}
\usepackage{amsmath}

\usepackage{listings}

\usepackage{hyperref}
\usepackage{xspace}

\usepackage{tabularx}

\usepackage{subcaption}

\usepackage{color}
\definecolor{bluekeywords}{rgb}{0.13,0.13,1}
\definecolor{greencomments}{rgb}{0,0.5,0}
\definecolor{redstrings}{rgb}{0.9,0,0}

\microtypecontext{spacing=nonfrench}

\ifisdraft
\newcommand{\redcomment}[1]{\textcolor{red}{\textbf{#1}} }

\else
\newcommand{\redcomment}[1]{ }

\fi

\newcommand{\NH}[1]{\redcomment{NH: #1}}
\newcommand{\DS}[1]{\redcomment{DS: #1}}

\ifisdraft
\usepackage{layouts}
\fi

\newcommand{\etal}{\emph{et al.}\xspace}

\newcommand{\ei}{{\it i)}\xspace}
\newcommand{\eii}{{\it ii)}\xspace}
\newcommand{\eiii}{{\it iii)}\xspace}

\usepackage{tikz}
\newcommand*\circled[1]{\tikz[baseline=(char.base)]{
            \node[shape=circle,draw,inner sep=1pt] (char) {#1};}}

\makeatletter
\newcommand{\customlabel}[2]{  \protected@write \@auxout {}{\string \newlabel {#1}{{#2}{\thepage}{#2}{#1}{}} 
  }  \hypertarget{#1}{#2}
}
\makeatother

\newcounter{invariant}
\newcommand{\invariant}[3] {
  \refstepcounter{invariant}
  \begin{center}
    \begin{small}
    \vspace{-1mm}
    \begin{tabular}{|p{0.95\columnwidth}|}
      \hline
      \textbf{\customlabel{inv:#1}{Invariant I\theinvariant}\ (#2)}
      #3
      \\\hline
    \end{tabular} 
    \vspace{-1mm}
    \end{small}
  \end{center}
}

\newcounter{assumption}

\newcounter{rights}
\newenvironment{rightsgroup}
  {\begin{center}\begin{small}  \vspace{-1mm}\begin{tabular}{|p{0.95\columnwidth}|}\hline}
  {\hline\end{tabular}\end{small}\vspace{-1mm}\end{center}}
\newcommand{\rightsgroupentry}[3] {
    \vspace{-0.6em}
   \refstepcounter{rights}\textbf{\customlabel{right:#1}{Right R\therights}\ 
   (#2)} #3\\}

\newcounter{requirement}

\usepackage{listings}

\DeclareCaptionFont{white}{\color{white}}

\lstdefinestyle{customc}{
  belowcaptionskip=0pt,
  aboveskip=5pt,
  belowskip=1.5pt,
  breaklines=true,
  xleftmargin=\parindent,
  language=C,
  morekeywords={uint64_t, uint32_t, uin16_t, uint8_t},  
  showstringspaces=false,
  basicstyle=\footnotesize\ttfamily,
  keywordstyle=\bfseries\color{green!40!black},
  commentstyle=\itshape\color{purple!40!black},
  identifierstyle=\color{blue},
  stringstyle=\color{orange},
  xleftmargin=0pt,
  framexleftmargin=0pt,
  framexrightmargin=0pt,
  framexbottommargin=1pt,
  framextopmargin=1pt,
  tabsize=2,
  captionpos=b,
}

\lstdefinestyle{customhaskell}{
  belowcaptionskip=0pt,
  breaklines=true,
  aboveskip=5pt,
  belowskip=1.5pt,
  language=Haskell,
  morekeywords={Natural,Name,retype,Set},  
  showstringspaces=false,
  captionpos=b,
  basicstyle=\linespread{0.9}\footnotesize\ttfamily,   keywordstyle=\bfseries\color{green!40!black},
  commentstyle=\itshape\color{purple!40!black},
  columns=fixed,
  basewidth={0.5em,0.45em},
  identifierstyle=\color{blue},
  stringstyle=\color{orange},
  xleftmargin=0pt,
  framexleftmargin=0pt,
  framexrightmargin=0pt,
  framexbottommargin=1pt,
  framextopmargin=1pt,
  tabsize=2
}

\lstdefinestyle{customisabelle}{
  basicstyle=\normalsize\normalfont,   belowcaptionskip=0pt,
  aboveskip=5pt,
  belowskip=1.5pt,
  breaklines=true,
  xleftmargin=\parindent,
  language=Haskell,
  morekeywords={record},  
  showstringspaces=false,
  captionpos=b,
  basicstyle=\footnotesize\ttfamily,
  keywordstyle=\bfseries\color{green!40!black},
  commentstyle=\itshape\color{purple!40!black},
  identifierstyle=\color{blue},
  stringstyle=\color{orange},
  xleftmargin=17pt,
  framexleftmargin=17pt,
  framexrightmargin=5pt,
  framexbottommargin=4pt,
  tabsize=2
}

\lstdefinestyle{customprolog}{
  basicstyle=\normalsize\normalfont,   belowcaptionskip=0pt,
  aboveskip=5pt,
  belowskip=2pt,
  breaklines=true,
  xleftmargin=\parindent,
  captionpos=b,
  language=Prolog,
  morekeywords={record},  
  showstringspaces=false,
  basicstyle=\linespread{0.9}\footnotesize\ttfamily,
  keywordstyle=\bfseries\color{green!40!black},
  commentstyle=\itshape\color{purple!40!black},
  identifierstyle=\color{blue},
  stringstyle=\color{orange},
  xleftmargin=0pt,
  framexleftmargin=0pt,
  framexrightmargin=0pt,
  framexbottommargin=1pt,
  framextopmargin=1pt,
  tabsize=2
}

\newcommand{\tright}[1]{\lstinline[breaklines=false,basicstyle=\ttfamily,identifierstyle=\sc\bfseries\color{green!40!black}]|#1|}
\newcommand{\toperation}[1]{\lstinline[breaklines=false,basicstyle=\ttfamily,identifierstyle=\color{blue}]|#1()|}
\newcommand{\tobject}[1]{\lstinline[breaklines=false,basicstyle=\ttfamily,identifierstyle=\sc\bfseries\color{red}]|#1|}

\newcommand{\grantauthority}{\tright{grant}\xspace}
\newcommand{\mapauthority}{\tright{map}\xspace}
\newcommand{\accessauthority}{\tright{access}\xspace}

\newcommand{\myparagraph}[1]{\noindent\textbf{#1.~~}\nolinebreak[4]}

\usepackage{enumitem}

\newenvironment{myitemize}{\begin{itemize}[itemsep=0pt,parsep=1pt,itemindent=8pt,labelsep=3pt,listparindent=0pt,topsep=0pt,leftmargin=0pt]}
{\end{itemize}}

\newenvironment{myenumerate}{\begin{enumerate}[itemsep=0pt,parsep=1pt,itemindent=10pt,listparindent=0pt,labelsep=3pt,topsep=1pt,leftmargin=0pt]}
{\end{enumerate}}

\ifisdraft

\usepackage[framemethod=default]{mdframed}

\mdfdefinestyle{inv}{	linecolor=black,linewidth=2pt,	topline=false,
	leftline=false,
	innertopmargin=1pt,
	skipbelow=\topskip, 	skipabove=\baselineskip,
}

\mdfdefinestyle{defi}{	linecolor=black,linewidth=2pt,	frametitlerule=true,	frametitlebackgroundcolor=gray!20,
	innertopmargin=3pt,
	skipbelow=\topskip, 	skipabove=\baselineskip,
	nobreak=true}

\mdtheorem[style=defi]{defi}{Definition}
\mdtheorem[style=defi]{op}{Operation}
\mdtheorem[style=inv]{inv}{Invariant}

\fi

\begin{document}

\date{}

\title{\Large \bf Secure Memory Management on Modern Hardware}

\author{Reto Achermann, Nora Hossle, Lukas Humbel, Daniel Schwyn, David Cock, Timothy Roscoe\\
Systems Group, Department of Computer Science, ETH Zurich}

\maketitle

\newcommand{\system}{\textit{Barrelfish/MAS}\xspace}

\begin{abstract}

Almost all modern hardware, from phone SoCs to high-end servers with
accelerators, contain memory translation and protection hardware like
IOMMUs, firewalls, and lookup tables which make it impossible to
reason about, and enforce protection and isolation based solely on the
processor's MMUs.  This has led to numerous bugs and security
vulnerabilities in today's system software.

In this paper we regain the ability to reason about and enforce access
control using the proven concept of a \emph{reference monitor}
mediating accesses to memory resources. We present a fine-grained,
realistic memory protection model that makes this traditional concept
applicable today, and bring system software in line with the
complexity of modern, heterogeneous hardware.

Our design is applicable to any operating system, regardless of
architecture.  We show that it not only enforces the integrity
properties of a system, but does so with no inherent performance
overhead and it is even amenable to automation through code generation
from trusted hardware specifications.

    \end{abstract}

\section{Introduction}
\label{sec:introduction}

Both new, fully-verified kernels and traditional
production-quality operating systems rely on a model of memory
addressing and protection so simple it is rarely remarked on: RAM and
devices reside at unique addresses in a single, shared physical
address space, and all cores have homogeneous memory management units
(MMUs) which translate virtual addresses into this single physical
address space.

The OS running on the platform then fulfills two roles: First, it manages
\emph{resource allocation}. Virtual memory makes multiplexing hardware easier
by decoupling the application's view of memory from the physical resources
managed by the OS, allowing \emph{late binding} of addresses.  Second it
forms, alongside the MMU, a \emph{reference
monitor}~\cite{Anderson:1972:reference_monitor}: All resource accesses
(dereferences) are intercepted by the monitor (specifically the TLB), and
checked against an \emph{access-control policy}.  This has for decades formed
the basis for secure process isolation in \emph{all} operating systems 
implementing virtual memory.

The reference monitor concept repeats throughout traditional OS design, with
more sophisticated abstractions gradually built up, and their associated
security properties enforced through a combination of hardware-provided
monitors (e.g.~MMUs), and software ones (e.g.~traps and syscalls).

For example, consider name (or address) resolution and
authorization checks in the \verb+mmap()+ syscall.  A process begins
with a \emph{reference} to a file: its filename.  The OS, meanwhile, enforces
some access-control policy, e.g.~UNIX-style permissions.  The calling process
\emph{dereferences} the filename by passing it to the \verb+open()+ syscall,
whereupon the OS validates the request against policy (permissions), and
\emph{resolves} the reference to another reference: the file descriptor (FD), now
referring to an entry in the global open-file table.  The existence of this
entry, and that the process may possess a reference is justified by the
top-level policy; The pattern of open files and FDs (the state) is a
\emph{projection} of something permitted by the policy.

This pattern is replicated in the VM system thanks to \verb+mmap()+.  Unix
cannot directly interpose on memory reads and writes (to the buffer cache page
mapped to the user), but does implement the initial \verb+mmap()+ call, and
the page fault handler.  The kernel builds a reference monitor by
\emph{composing} itself with that provided by the MMU.  On an \verb+mmap()+
call, the kernel verifies that the FD is valid, with appropriate permissions
(e.g.~write), before constructing a VM region to back the mapping.  The policy
encoded in the region's flags is thus a (transitive) projection of the
original file system permissions.  On a page fault, the kernel is again invoked 
to lazily populate the region (from the buffer cache). Now, it can consult the
mapping parameters (e.g.~writable), and translate these to flags in the
page-table entry.

Thus, the page-table state (e.g.~permission bits), and thence the
eventual TLB state, are justified by a chain of monitors all the way
back up to the system policy (file system permissions).  The MMU
enforces this \emph{projected} policy on the OS' behalf.  Together
they form, in security terms, a \emph{compound reference monitor} to
enforce a policy both on real hardware resources (RAM), and abstract
OS-specific objects (processes, files).

This model has worked well for decades, but has been undermined by a
changing hardware contract.  A modern system contains not just 
processors and their attached MMUs, but system MMUs or IOMMUs, memory firewalls, 
region lookup tables, etc. all of which mediate access to and from parts of the
platform.  ``Smart'' devices like GPGPUs, co-processors, network cards, or 
accelerators come with their own hardware protection and translation
units~\cite{Gerber:2015:YPP}. 

In such a system, the processor's MMU alone does not form a reference monitor
for memory, as it is not invoked on all accesses.  Indeed, the
complex address-translation topology of these systems renders even the concept
of a unique physical address meaningless, raising the risk that the policy
encoded into the distributed hardware reference monitor (the collections of
MMUs, SMMUs, etc.) is inconsistent due to their differing views of the
machine.  These two problems have already led to security
vulnerabilities~\cite{Markettos:2019:TEV, Markuze:2016:TIP, CVE-2013-4329,
CVE-2015-6994}.

We identify three classes of security vulnerabilities and bugs
(\autoref{tab:bugseliminated}) that \ei cause the execution of an operation
without sufficient rights (a failure of \emph{policy enforcement}), \eii allow
a compromise of the reference monitor itself (e.g.~writing translation tables,
a failure of \emph{partitioning}), or \eiii use the wrong addresses in
descriptors or pointers (a failure of \emph{name resolution}). The lack of a
proper reference monitor which is aware of the complex and configurable
addressing network continues to result in numerous bugs and security
vulnerabilities~\cite{CVE-2015-4421,CVE-2015-4422, CVE-2016-5349,
Chester:2018:ECV, Gong:2019:EQW, Schnarz:2014:TAR, Zhu:2017:USD}.  Confining
these bugs in a kernel is hard, and they are likely to compromise the entire
system~\cite{Bigs:2018:JMOS}.

In this paper we demonstrate that these whole classes of bugs can be prevented
by extending the traditional OS-MMU reference monitor to cover \emph{all} 
hardware translation and enforcement engines, allowing policy enforcement on all 
memory accesses, ensuring consistent name resolution by adopting the 
\emph{decoding net}~\cite{Achermann:2017:FMA,Achermann:2018:PAR} as a more 
faithful model of modern addressing hardware, and ensuring the secure 
partitioning of reference monitor state either through a partitioned capability 
system, or in a traditional kernel (such as Linux) by good software engineering 
practice and the application of existing memory management interfaces.

Our first contribution is to identify the undermining of the traditional
OS-MMU reference monitor by a changing hardware/software contract as 
the root cause of several large classes of critical security bugs.

Our second contribution is to adopt a faithful model of complex addressing
hardware (the decoding net), and from it derive a minimal
\emph{least-privilege} model of memory management authority on modern
hardware, covering the common functionality of all virtual memory systems
(\autoref{sec:model}).

Our third contribution is the specification of an OS-agnostic \emph{reference
monitor} to enforce policy expressed in the above model, prototyped as an
\emph{executable specification} in Haskell, and abstracting the OS's internal
policy language (e.g.~capabilities or ACLs) as an \emph{access-control matrix}.

Our fourth contribution is to demonstrate that this reference monitor design
can be implemented without invasive changes on either partitioned capability
systems (e.g.~seL4 or Barrelfish), or on ACL-based UNIX-style kernel (such as
Linux).  Further our benchmarks demonstrates that there is no measurable
performance cost for a secure fully-explicit least-privilege system-wide virtual 
memory authority implementation (\autoref{sec:evaluation})

\section{Eliminating Classes of Bugs}
\label{sec:motivation:bugpotential}

\begin{table}
  \begin{footnotesize}
  \begin{tabular}{lp{0.60\columnwidth}}
    \textbf{Type} & \textbf{CVE-...} \\
    Policy enforcement & 1999-1166 2014-3601 2014-8369 2014-9888 2017-16994  2019-2250 
    2019-10538 2019-10539 2019-10540 \\
    Partitioning & 2011-1898 2013-43292014-0972 2018-1038 2018-11994 2019-2182 
    2019-19579\\
    Name resolution & 2013-4329 2014-9932 2016-3960 2016-5349 2017-8061 
    2017-12188 2019-15099
  \end{tabular}
  \end{footnotesize}
  \caption{Classes of Security Vulnerabilities.}
  \label{tab:bugseliminated}
\end{table}

The difficulty of getting complex memory addressing right in an OS is
shown by the steady, ongoing stream of related bugs and
vulnerabilities in operating systems, for example, policy enforcement
in Linux's memory management code~\cite{Huang:2016:ESL}.

We identify three classes of common bugs and security vulnerabilities related
specifically to the incompleteness of the current reference monitor, which
would be rendered impossible under comprehensive reference monitor
which faithfully reflected the hardware:

\myparagraph{Policy Enforcement}
These are bugs where a subject was able to change 
the configuration of a translation unit without having the proper rights do to 
so. The reference monitor fails here to enforce the system policy:
\begin{itemize}
\item Mappings with holes belonging to another subject~\cite{CVE-2014-3601}.
\item Incorrect permissions on data pages~\cite{CVE-2014-9888}.
\item IOMMU configured to map too large a
range~\cite{CVE-2019-10538,CVE-2019-10539,CVE-2019-10540}.
\end{itemize}
All these bugs are impossible once the operations are performed through a
(correct) reference monitor implementing the system security property.

\myparagraph{Partitioning}
These bugs involve bypassing the reference monitor directly e.g.~by directly
modifying its internal state:
\begin{itemize}
\item DMA transfers into MSI-x interrupt registers~\cite{CVE-2011-1898}.
\item DMA transfers into IOMMU control registers~\cite{CVE-2014-0972}.
\item Process modifies its own page table~\cite{CVE-2018-1038}.
\end{itemize}
These are prevented once the reference monitor state is identified and
\emph{partitioned} by subjecting them to system policy e.g.~that no DMA engine
or process may map a page table.

\myparagraph{Name Resolution} 
This class represents inconsistent interpretations of pointers (names):
\begin{itemize}
\item Insufficient context to identify the correct
object~\cite{CVE-2016-5349}.
\item Resolving addresses in the wrong context~\cite{CVE-2017-12188}.
\end{itemize}
These are prevented once names are dereferenced (resolve) through a monitor
with a complete, accurate model of addressing.

\section{Background and Related Work}
\label{sec:background}

Before presenting our authority model and the executable specification in the
next section, we will briefly cover reference monitors in a little more
detail, in particular the importance of consistent naming, and how complex
addressing topologies make it difficult.

We also summarize the existing decoding net model, the executable
specification/refinement approach which we borrow from the seL4 system, and the related work.

\subsection{Reference Monitors}

The reference monitor is a powerful
structuring concept in access control, and is implicitly used in practically
every OS.  A reference monitor enforces an access-control policy, allowing a
separation of concerns, and thus effort: if \emph{every} access is subject to
the policy, then the overall safety of the system (w.r.t.~the policy) can be
guaranteed \emph{independently} of the correctness of the components making
the accesses.  This is of enormous benefit to a monolithic system (e.g.~Linux), 
where a fault in one subsystem can easily spread to others,
particularly as any subsystem can, in principle modify translations.  Even
without enforcing a strict boundary between components (as in a microkernel),
routing all updates via a single component responsible for safety ensures that
\emph{accidental} errors will no longer lead to a whole-system compromise.

The critical point for a reference monitor is that \emph{all} accesses
must pass through to it, and that it is able to accurately identify which resources are
being accessed (e.g.~which DRAM address will ultimately be written) when
applying its policy.  Both of these are undermined in the complex
address-translation networks of modern systems, but not fatally so: The hardware
component of the reference monitor is now \emph{distributed} among multiple
system MMUs, firewalls, etc.; addresses may be rewritten \emph{after} policy is
applied, routing them to locations that should not be accessible.

Both of these problems are solved with an accurate model of the hardware:
First, to know the complete set of access-control components that must be
included in the reference monitor, and second, to guarantee that any
translation below the access-control level is consistent with policy.

\subsection{The Canonical Name Problem}

As established, modern platforms are composed of multiple, heterogeneous cores
and devices each of which can issue accesses to addressable resources such as
DRAM, non-volatile memory or device registers.  Worse, there is no single
``reference'' physical address space~\cite{Gerber:2015:YPP}.  Instead, a
network of address spaces or buses is connected by address translation units
which ``route'' memory accesses.  As just described, in order to securely
enforce access control, it is essential to know what final resource some
intermediate address (or \emph{name}) refers to.

I/O memory management units (IOMMUs, or system MMUs) translate addresses
generated by accelerators and DMA-capable devices into a ``canonical''
system-wide physical address space.  This allows user-space programs to share
a virtual address space with a context on the device, but impose a further
complexity burden on the underlying OS which must now ensure that IOMMUs are
always correctly programmed.  This code is fraught with complexity and
consequent bugs and vulnerabilities, as it is also intended to provide
protection from malicious memory accesses~\cite{Morgan:2016:BIP,
Morgan:2018:IPIO, Markuze:2016:TIP, Markettos:2019:TEV}.  The problem is
likely going to get worse with the proliferation of IOMMU designs built into
GPUs, co-processors, and intelligent NICs. 

Even memory controllers can violate the traditional model.  Hillenbrand
\etal~\cite{Hillenbrand:2017:MPM} reconfigure memory controller configurations
from system software to provide DRAM aliases for mitigating the performance
effects of channel and bank interleaving.  Proposals for ``in-memory'' or
``near-data'' processing~\cite{Patterson:1997:CIR, Vermij:2017:AIN,
Zhang:2014:TTP} raise further questions for OS 
abstractions~\cite{Barbalace:2017:TTO} and require a way to unambiguously
refer to memory regardless of which module accesses it.

\subsection{Decoding Nets}

A systematic and accurate way to establish canonical names for
access-controlled resources that may be referred by different \emph{local}
names in different parts of the system is provided by the established
\emph{decoding net}~\cite{Achermann:2017:FMA,Achermann:2018:PAR} model of
address translation.

Decoding nets model the addressing structure of a system as a directed graph,
where nodes represent (virtual or physical) address spaces or devices
(including RAM), and edges the translation of \emph{AS-local} addresses into
other address spaces or devices.  The graph is a set of nodes, defined as an
abstract datatype:
\begin{small}
\setlength{\abovedisplayskip}{3pt}
\setlength{\belowdisplayskip}{3pt}
\begin{align*}  
\textit{name} &= \textrm{Name}\ \textit{nodeid}\ \textit{address}\\
\textit{node} &= \textrm{Node}\ \textbf{accept}::\{\textit{address}\}\  \\
               &\phantom{{}=\textrm{Node}\ }\textbf{translate} ::
                                \textit{address}\rightarrow \{name\}
\end{align*}
\end{small}
The model distinguishes \emph{local} names ($\textit{address}$), relative to
some address space, and \emph{global} names ($\textit{name}$), which qualify a
local name with its enclosing address space.  Each node may \textbf{accept} a
set of (local) addresses (e.g.~RAM or memory mapped device registers), and/or
\textbf{translate} them to one or more global names (addresses in other
address spaces, e.g.~MMU or PCI bridges).

This approach dovetails nicely with the reference monitor concept as described
above.  Every \textbf{translate} step corresponds to a \emph{dereference}
operation, and any \textbf{accept} can be used as a canonical name: the ID of
the accepting node, plus the \emph{local} address at which it accepts 
(e.g.~address within a DRAM bank).

Decoding nets have been successfully used to model a wide variety of systems
of exactly the sort that is of interest to us, and give a trustworthy, precise 
guide
to where a reference monitor is required: any \emph{configurable} translation
node must be treated as part of the distributed reference monitor.  It must
only be configured such that its local translations are a \emph{projection} of
the higher-level security property, exactly as for a processor's MMU.
\emph{Static} configuration nodes must be configured in such a way (either by
construction or static verification) that their translations are consistent
with the projected policy at the point they are applied.

\subsection{Refinements and Executable Specifications}

\begin{figure}
\includegraphics{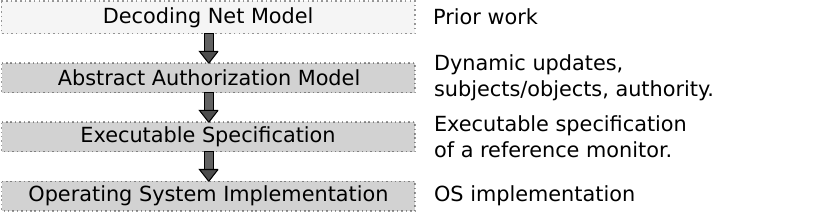}
\caption{Methodology Overview: Refinement steps.}
\label{fig:methodology:overview}
\end{figure}

We borrow our modeling technique, combining \emph{refinement} with
\emph{executable specification} from the successful seL4 project.  We identify
all relevant \emph{objects} (page tables, address spaces, frames, \ldots), the
\emph{subjects} that manipulate them (processes, devices, \ldots), and which
\emph{authority} each subject exercises over each object (e.g.~in mapping a
frame to a virtual address).  These are expressed in an \emph{access-control
matrix} (following Lampson~\cite{Lampson:1974:Protection}) which forms our
\emph{abstract specification}, analogous to the high-level \emph{security
policy} (integrity) shown to be refined (correctly implemented) all the way
down to compiled binaries for seL4~\cite{Sewell:2011:Integrity}.

Again, as in seL4~\cite{Cock:2008:SMS}, we next develop an executable
specification in Haskell (see \autoref{sec:haskell}), expressing subjects,
objects, and authority as first-class objects, permitting rapid prototyping
without giving up strong formal semantics.  Correspondence between abstract
and executable models is thus far by inspection and careful construction.

Finally, we show (again with precedent~\cite{Winwood:2009:MG}) that the
executable model (and hence the abstract model) permits multiple
high-performance implementations (see \autoref{sec:implementation}): On
Barrelfish, as a representative of partitioned-capability systems including seL4
(capabilities corresponding to \emph{rows} in the matrix), and on Linux, as a
representative UNIX-style monolithic kernel (where ACLs correspond to
\emph{columns} in the matrix).

\subsection{Related Work}

The seL4 proof~\cite{Klein:2009:SFV} assumed a single, fixed, physical address
space and a single MMU, and provides no guarantees in the presence of other
cores or DMA devices.  CertiKOS~\cite{Gu:2016:CEA} builds on a model of memory
accesses to abstract regions of private, shared or atomic memory, but again
provides no proof in the presence of other translation units or cores.  Even
work on verifying memory consistency in the presence of translation currently
only considers the simple case of virtual-to-physical
mappings~\cite{Romanescu:2010:SDV}.

Graviton~\cite{Volos:2018:GTE} provides a trusted execution environment for
GPUs requiring all updates to the page tables go through the command
processor, acting as a reference monitor for the GPU.
Komodo~\cite{Ferraiuolo:2017:KUV} uses ARM TrustZone~\cite{man:arm:trustzone}
to implement a software enclave.  Both of these works are steps in the right
direction, and in this work we extend this approach to the whole system.

OpenCL's Shared Virtual Memory~\cite{man:opencl}, nVidia's
CUDA~\cite{man:nvidia:UMCUDA} or HSA~\cite{man:hsa} provide a unified view of
memory, ensuring addresses remain valid between CPU and GPU.
VAST~\cite{Lee:2014:VIL} which uses compiler support to dynamically copy
memory to and from the GPU and Mosaic~\cite{Ausavarungnirun:2017:MGM}, which
provides support for multiple sizes of page translation in a shared virtual
address space between CPU and GPU.  These approaches ensure address
consistency in the specific case of CPU--GPU sharing, but are again not
whole-system approaches.

In DVMT~\cite{Alam:2017:DVM}, a customized TLB miss handler implemented as a
helper thread installs entries in the TLB using specialized instructions.
Similar to the MMU, the OS/hypervisor sets up data structures specifying the
policy which mappings the thread is allowed to install. Again this solution
focuses on the processor and its MMU.

\section{Model}
\label{sec:design}

A static \emph{decoding net} is a snapshot of the address translation
configuration of a system, at a particular moment.  We augment the static
decoding net with a transition relation, modelling the dynamic reconfiguration
of the translation hardware such as when a page table is modified.  The
allowable transitions express the actions (or \emph{traces}) permitted by the
model.

\subsection{Authority and Dynamic Behavior}
\label{sec:model}

\begin{figure}
  \begin{center}
  \begin{footnotesize}
    \begin{tabular}{|p{0.95\columnwidth}|}\hline
\textbf{Nodes:} $\textit{node} :: \textit{Decoding Net Node}$\\
\textbf{Objects:}  
$\textit{Object} = \{\textit{name}\}$\\
\textbf{Rights:}  
$\textit{Right} = \ \textit{Grant}\ |\ \textit{Map}|\ \textit{Access}$ \\

\textbf{Configuration Space:} \\
$\texttt{ConfSpace} :: \textit{AddressSpace} \rightarrow \{\textit{node}\}$\\

\textbf{Address Space Configuration:} \\
$\texttt{Configuration} :: \textit{AddressSpace} \rightarrow \textit{node}$\\

\textbf{Access Control Matrix:}\\
$\texttt{AccessControlMatrix} :: \textit{Subject} \times \textit{Object} 
\rightarrow 
\{\textit{Right}\}$

\textbf{Model State:}\\
$State = (\texttt{AccessControlMatrix}, \texttt{Configuration})$
\\
\textbf{State Transitions:}\\
$ \texttt{ModifyMap} :: \textit{Subject} \rightarrow 
                        (\textit{name} \rightarrow
                        \{\textit{name}\}) \rightarrow 
                        \textit{State} \rightarrow
                        \textit{State}
$
    \\\hline\end{tabular} 
\end{footnotesize}
    \end{center}
  \vspace{-4mm}
  \caption{Model Definition}
  \label{fig:model:definition}
\end{figure}

The system consists of a set of \emph{address spaces} each having a current
\emph{configuration}, which corresponds to a \emph{decoding net} node, that
defines the translation of local addresses in this address-space
\emph{context}:
\setlength{\abovedisplayskip}{3pt}
\setlength{\belowdisplayskip}{3pt}
\begin{displaymath}
\textit{configuration} \mathop{::} \textit{address space} \rightarrow 
\textit{node}
\end{displaymath}
This lets us reason about translations with the existing mechanisms available
for decoding nets. Hardware constraints, e.g.~an MMU that only supports the
translation of naturally aligned 4 KiB blocks of addresses, are expressed as a
restriction on the set of possible nodes an address space can map to.  This
set is the \emph{configuration space} of an address space.
\autoref{inv:wellformed} requires that every address space must have a
well-defined configuration. The configuration space of a fixed address space
is a singleton set.

\invariant{wellformed}{Well-defined Configuration}{\\
  $\forall a::\textit{AddressSpace}.\ \texttt{Configuration}\ a \in 
\texttt{ConfSpace}\ a.$}

\begin{figure}
  \includegraphics{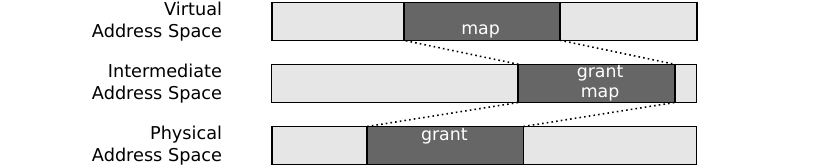}
  \caption{Mappings between 
  address spaces showing grant and map rights of mapped segments.}
  \label{fig:model:single-address-space}
\end{figure}

\myparagraph{Configuration Authority (Mapping)}
The configuration of some address spaces can be changed. The configuration
space defines the set of \emph{possible} states an address space may occupy.
An \emph{authority} is a subset of configuration transitions, representing
what configuration actions a given subject is permitted to take.

Consider~\autoref{fig:model:single-address-space}, representing the general
case of an update to an intermediate address space (for example the
intermediate physical address, IPA, in a two-stage translation system).  We
identify two distinct authorities: The \mapauthority authority, or the
authority to change the meaning of an IPA by changing its mapping; and the
\grantauthority authority, or the right to grant \accessauthority (by mapping)
to some range of physical addresses. Note that the 'virtual' and 'physical'
address spaces of~\autoref{fig:model:single-address-space} can be viewed as
special cases of an intermediate address space: A top-level 'virtual' address
space is simply one to which \emph{nobody} has a \grantauthority authority,
and a 'physical' address space e.g.~DRAM is one to which there exists no
\mapauthority authority.

\begin{rightsgroup}
  \rightsgroupentry{grant}{Grant}{\\
  The right to insert \emph{this} memory object into \emph{some} address space}
  \rightsgroupentry{map}{Map}{\\
  The right to insert \emph{some} memory object into \emph{this} address space}
  \rightsgroupentry{access}{Access}{\\
  The right to read or write an object.}
\end{rightsgroup}

\begin{figure}
  \includegraphics{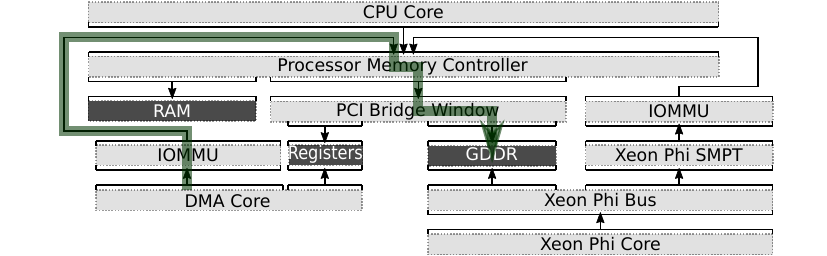}
  \caption{Address spaces in a system with two PCI devices}
  \label{fig:model:multiple-address-spaces}
\end{figure}

\myparagraph{Changing Mappings}
Consider~\autoref{fig:model:multiple-address-spaces}, showing the address
space configuration of a system with two PCI devices: a DMA engine and an
Intel Xeon Phi co-processor.  Imagine that we wish to establish a shared
mapping to allow a process on a Xeon Phi core to receive DMA transfers
(e.g.~network packets) into a buffer allocated on the GDDR (following the
highlighted path from the DMA core to the GDDR).

The process `owns' the buffer, and has the ability to call \toperation{recv},
triggering a DMA transfer.  In other words, the process has the right to grant
\accessauthority (temporarily) to the DMA core, but it clearly should not have
the ability to modify the IOMMU mappings of the DMA core at will.  Hence, it
does not have the \mapauthority authority on the relevant address space.

To change the mappings of an address space, an agent (a \emph{subject}, in
standard access-control terminology) needs both the \grantauthority authority
on the buffer \emph{object}, and the \mapauthority authority on the address
space \emph{object}.  

The state transition, i.e.~changing the \emph{configuration} and therefore how
an address space translates addresses, is expressed by the operation
\toperation{ModifyMap}: A subject tries to change how a name is being
translated by the system, and thus updates its state.

\begin{table}
  \begin{footnotesize}
  \begin{center}
    \begin{tabular}{l|ccc}
      \emph{subject} / \emph{object}  & DMA IOMMU    & buffer \\
      \hline \\
      IOMMU driver                  & \mapauthority &        \\
      Xeon Phi process              &              & \grantauthority
    \end{tabular}
  \end{center}
\end{footnotesize}
  \vspace{-4mm}
  \caption{Access control matrix of the Xeon Phi example}
  \label{tab:model:acm}
\end{table}

\myparagraph{Authority Representation} 
In a monolithic kernel, both these authorities are held
(implicitly) by the kernel, which exercises them on behalf of the subjects.
It is up to the kernel to maintain accurate bookkeeping to determine whether
any such request is safe, typically using an ACL (access-control list)
i.e.~the \emph{object} lists the subjects and their authorities on it.
In a partitioned-capability system such as seL4 or Barrelfish, these
authorities are represented by capabilities, handed explicitly to one
\emph{subject}, to authorize the operation.  In this case, subjects hold the
authority on the \emph{object}.  These are equivalent from the perspective of
access control, differing only in implementation: the same two basic types of
authority are present.

The standard representation of authority in systems is an access control
matrix~\cite{Lampson:1974:Protection}, such as that
of~\autoref{tab:model:acm}.  This can be read in rows: The IOMMU driver has
the \mapauthority \emph{capability} to the IOMMU address space, and the
process the \grantauthority capability to the buffer.  Alternatively, reading
down the columns gives the ACLs: the IOMMU records \mapauthority
\emph{permission} for the driver, and for the buffer records a \grantauthority
\emph{permission} for the process.

\myparagraph{Security Property}
This access control matrix is our abstract model.  A system
is correct (secure) \emph{statically}, if its current configuration is
consistent with the access control matrix.  It is secure \emph{dynamically} if
any possible transition, beginning in a secure state, must leave the system in
a secure state. The access control matrix, together with the configuration space 
defines the allowable state transitions. The address space must have a valid
configuration supported by hardware, and the subject modifying it must have
sufficient rights to do so.

\subsection{Executable Specification}
\label{sec:haskell}

We refine this abstract model into an executable specification of a
\emph{reference monitor}~\cite{Anderson:1972:reference_monitor} for
\toperation{ModifyMap}.  When composed with the reference monitor
\accessauthority i.e.~the MMU, we have our desired compound reference monitor
for the fully-dynamic VM system, secure for accesses beginning at any core or
device.

This specification serves as an intermediate step between
(\autoref{fig:methodology:overview}) the abstract model and the concrete OS
implementation of the next section, and also an OS-agnostic prototype for
implementation in other systems.  This approach is inspired by
\emph{seL4}~\cite{Derrin:2006:RMA}, which also employed an intermediate
Haskell specification to facilitate prototyping.

\myparagraph{Explicit Translation Structures}
We now explicitly represent address translation structures (e.g.~page tables,
or memory-mapped device registers) as memory objects, without imposing any
particular layout on them. This allows us to reason about the manner in with
address translation depends on the contents of a memory object (e.g.~page
tables in RAM, or the contents of device registers).

Once the translation structures are explicit, and noting that these are
exactly the reference monitor state we must securely partition, we can state
the partitioning invariant (\autoref{inv:obj:noaccess}) in terms of
implementation-visible objects.

\invariant{obj:noaccess}{Partitioning}{\\
No subject has \accessauthority to a translation object}

\begin{figure}
  \begin{center}
    \includegraphics{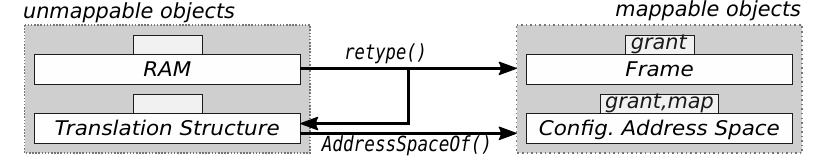}
    \caption{Object Type Hierarchy and possible rights.}
    \label{fig:objecttypes}
  \end{center}
\end{figure}

We model address translation structures as an opaque data type
(\lstinline|TStructure|). This allows us to maintain generality by assuming
nothing about their actual inner structure:

\begin{lstlisting}[style=customhaskell]
data Object = RAM {base::Name, size::Natural} 
            | Frame {base::Name, size::Natural} 
            | TStructure {base::Name, size::Natural} 
\end{lstlisting}

\noindent
Memory objects form a hierarchy (\autoref{fig:objecttypes} shows an excerpt) 
which defines how the different types of objects can be \emph{derived} from each 
other.  For example, in-memory translation structures (\tobject{TStructure}) are 
created by retyping \tobject{RAM} objects. \tobject{RAM} is the base type for 
\emph{untyped} memory. Retyping \tobject{RAM} to a \tobject{Frame} makes it 
possible to map it into an address space i.e.~to \grantauthority access to it.
Note, that neither \tobject{RAM} nor \tobject{TStructure} have the 
\grantauthority right, and therefore these may never become accessible
(partitioning).

An address space is derived from (and defined by) a translation structure, and
is an explicit object granting the right to map this space into higher-level
address spaces (e.g. a second-stage page table defining an IPA space, assigned
to the guest-physical address space of a virtualized OS):
\autoref{fig:model:single-address-space}.

\begin{lstlisting}[style=customhaskell]
AddressSpaceOf :: Object -> AddressSpace
\end{lstlisting}

\myparagraph{Authority and State}
The system is a set of agents, a \emph{mapping database} (MDB) recording the
derivation relation between objects, and a set of active address spaces:

\begin{lstlisting}[style=customhaskell]
data KState = KState (Set Agent) MDB (Set AddrSpace) 
\end{lstlisting}
Authority is either directly to an object, or a meta-authority, the right
to grant an authority to another.  In turn set of such authorities, coupled
with an identifier, define an agent.
\begin{lstlisting}[style=customhaskell]
data Authority = Access Object | Map Object 
               | Grant Authority
\end{lstlisting}

\myparagraph{Reference monitor}
The model exposes a set of operations that either change a configuration or
access a memory address. The set of permitted operations defines the behavior
of the reference monitor.  We express this in Haskell as a custom state monad: 
\begin{lstlisting}[style=customhaskell]
data Operation a = Operation (State -> (a, State))
  instance Monad (Operation) where ...
\end{lstlisting}
The reference monitor intercepts operations and verifies that the agent
performing the operation has sufficient rights to execute it.  We express the
changes to the system's state as sequence of operations on the reference monitor,
e.g.~\toperation{retype} or \toperation{map}, forming a trace of operations:

\begin{lstlisting}[style=customhaskell]
mappingTrace = do
    ...
    -- retype a RAM object to a Frame    
    res <- Model.retype RAM Agent Frame Agent
    -- retype another RAM object to a TStructure
    res <- Model.retype RAM2 Agent TStructure Agent
    -- map the frame into the translation structure
    mapping1 <- Model.map TStructure Frame Agent
    ...
\end{lstlisting}
Model traces are sequences of monitor states, (\tobject{KState}), each
corresponding to a static decoding net model. Operations include:
\begin{myitemize}
    \item \toperation{retype} converts an existing object into an object of a
    permissible sub type.
    \item \toperation{map} installs a mapping in a translation structure.
    \item \toperation{copy} copies an authority from one subject to another.
\end{myitemize}

\myparagraph{Valid Traces} Contained within the set $T$ of all possible
traces, there is a set of traces $T_V \in T$ that conform to all constraints
enforced by the executable specification. We express these traces in the model
as sequences of \lstinline|KState|s. All other traces ($T - T_V$) indicate
ending in a failure state (e.g.~that execution ended in a state not satisfying
the access-control policy).

\myparagraph{Summary} The executable specification allows us to both simulate
and specify sequences of operations such as memory accesses or translation
configurations as they would be performed by a concrete OS, implementing the
new abstract model.

\section{Implementation in a real OS}
\label{sec:implementation}

In this section, we describe the implementation of the reference monitor and
runtime support libraries and services in two classes of operating systems: a
complete implementation in \system\footnote{MAS stands for \textbf{m}ultiple
\textbf{a}ddress \textbf{s}paces.} as a representative of a partitoned
capability system, derived from the open-source Barrelfish
OS~\cite{Baumann:2009:MNO}, and side-by-side a sketch of an implementation
within Linux, as a representative of a traditional UNIX-style kernel. 

\begin{figure}
  \includegraphics[width=\columnwidth]{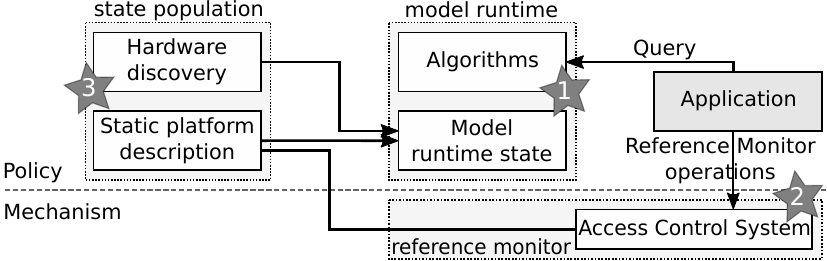}
  \caption{Implementation Overview}
  \label{fig:impl:overview}
\end{figure}

\myparagraph{Architecture Overview}
\autoref{fig:impl:overview} shows an overview of the resulting architecture.
We separate policy and mechanism: \circled{1} at the center is the runtime
representation of the model (\autoref{sec:impl:runtime}) which stores the
memory topology and provides queries and algorithms for memory allocation
policies, \circled{2} the reference monitor which enforces access control and
provides the mechanisms for resource management and configuration, and
\circled{3} static platform descriptions and dynamic discovery mechanisms
(\autoref{sec:impl:modelpopulation}) provide input for the policy and
mechanism implementations.

\subsection{Runtime Support}
\label{sec:impl:runtime} 

We implement the runtime representation of the address space model
(\autoref{fig:impl:overview}, \circled{1}) in a policy engine. On Barrelfish,
this is merged into the Prolog-based system knowledge-base
(SKB)~\cite{Schupbach:2011:DLA}, which already stores both static and dynamic
facts about the system.  On Linux, we could use a standalone Prolog instance
and run it as a service, or implement the model directly along with other
memory allocation policies inside the kernel.  We now describe the model
representation, its algorithms and potential optimizations. 

\begin{lstlisting}[style=customprolog,caption=Prolog Model 
Representation,label=lst:impl:model,float]
assert(translate(RegionFrom,RegionTo)).
assert(overlay(NodeFrom,NodeTo)).
assert(accept(Region)).

dn_get_allocation_range(NodeSrc, NodeDst).
dn_get_config_nodes(NodeSrc, NodeDst).
dn_resolve_range(Node, Addr, Size).
dn_resolve_range(NodeSrc, Addr, Size, DstSrc).
\end{lstlisting}

\myparagraph{Model representation}
We implement the model representation by asserting facts for the accept,
translate and overlay constructs of the model (see syntax in
\cite{Achermann:2018:PAR}).  \autoref{lst:impl:model} shows the corresponding
Prolog rules. This encodes the decoding net, and adds the information to the
database.

\myparagraph{Algorithms} 
On top of the model encoding, we implement several algorithms, useful for
making allocation and configuration policy decisions. For instance, to set up
a device, the driver uses the \toperation{dn_get_allocation_range} query to
find a suitable address space for memory allocation, then runs
\toperation{dn_get_config_nodes} to get the list of address spaces which need
to be configured to make the memory resource accessible, and lastly execute
\toperation{dn_resolve_range} to obtain the address at which the device sees
the memory resource. 

The result of the queries is then converted into a sequence of capability 
operations to allocate memory, setup translation structures and perform the 
relevant mappings. Note, the model queries only provide a roadmap, the actual 
reconfiguration steps are invocations of the reference monitor which  
enforces the authority and integrity of the system following the 
definition of the executable specification (\autoref{sec:haskell}). 

\myparagraph{Optimization}
Running the Prolog queries on the full graph is costly. We provide a library
that caches the (flattened) graph representation consisting only of
cores/devices, configurable address spaces and memory nodes in the Prolog
engine \emph{and} directly in C using adjacency lists.  We can then run a
shortest-path algorithm to perform the queries, which minimizes the number of
address spaces to configure.

\subsection{Reference Monitor}
\label{sec:impl:refmon}

We now describe the implementation of the reference monitor defined by the executable 
specification in Linux and \system.

\myparagraph{Resource Management}
Both, Linux and Barrelfish already have thorough resource management
mechanisms, albeit diferent: Barrelfish manages physical resources using a
distributed, partitioned capability system for naming, access control, and
accounting of objects.  As in seL4~\cite{Elkaduwe:2008:VPM}, capabilities are
\emph{typed} to indicate what can be done with the memory they refer to; rules
dictate valid \emph{retype} operations (e.g RAM to a Frame). Linux maintains a
data structure, the page struct, for every 4 KiB page of memory. In both
systems, only the kernel has direct access to those data structures, and can
maintain the partitioning invariant.

\myparagraph{Reference Monitor}
As with all microkernels, Barrelfish's kernel is essentially nothing but a
reference monitor.  It uses the capability system to express the objects in
memory and the authority a process (subject) has over them. Any changes to the
translation units (e.g.~mapping a memory frame into the IOMMU) correspond to
capability operations. The reference monitor checks type, address spaces and
rights of the capabilites. \DS{Should we mention the drivers as "sub reference
monitors" here?}

On Linux, we can use the para-virtualization interface (PV-Ops) to implement a 
reference monitor inside the kernel itself. We can then extend the PV-Ops interface
to include all address translation units in the system. This effectively implements
a well-defined hypercall interface to request changes to the translation tables from 
the hypervisor acting as the reference monitor. Similarly, the nested 
kernel~\cite{Dautenhahn:2015:NKO} integrates a privileged kernel inside the monolithic 
kernel which interposes all updates to translation tables. Extending this interface to 
include all other translation hardware as well, would present a good way to implement 
a reference monitor inside the Linux kernel.
\DS{Are these two different approaches or do they complement each other?}

\myparagraph{Naming of Resources}
Barrelfish's capabilities contain physical addresses to identify the objects
they are referring to. To be able to still identify the objects uniquely in
the presence of multiple address spaces we change the capability system in
\system to use canonical base names, consisting of an address space identifier
and an address within that address space. We adapt the kernel to consider the
ASID when performing capability operations. An operation may now fail in new
ways, due to incompatible address spaces of the capabilities (e.g.~one cannot
directly map host physical frame to a guest virtual address). 

Linux uses the physical frame number (PFN) uniquely identify every 4 KiB page of
memory. Using the sparse memory model~\cite{lwn:2005:linuxsparsemem} or
heterogeneous memory~\cite{man:linux:hmm}, we can implement memory nodes
(address spaces) a dynamic mapping of the PFN to the underlying page struct.
In this manner, we can use the PFN as the memory resource's canonical name. 

On both operating systems, we need a function to dereference the canonical name of a 
resource into a locally valid address. We can \emph{generate} such a translation 
function based on the platform description or the model state.

\myparagraph{Object Types}
In addition, \system introduces new capability types for all hardware 
translation units (not just page tables), ASID allocation, and entire physical, 
intermediate or virtual address spaces. Like Barrelfish, we allow a capability 
to refer to a memory region of arbitrary size, but require that it must not span 
multiple address spaces. 

On Linux, we do not need to use typed objects as such as the kernel does not expose
handles to physical resources to user space. Internally, Linux already uses different
accounting types for memory allocations. 

\myparagraph{Page Tables and Address Spaces}
\system introduces distinct capability types for all hardware-defined 
translation structures  (register sets or page table levels). Each of these 
capability types are translation structures in the sense of the executable spec. 
Since a page table defines an address space, we can \emph{derive} an 
\textit{address space} capability from it, and use it to install
mappings in other address spaces. Deleting the page table capability triggers
a recursive deletion of its spanned address spaces and all possible mappings. 
We integrated this process into the capability system. This is effectively 
equivalent to \emph{revoking} all descendants of the address space capability 
and then deleting it. This ensures, that there are no mappings referring to an 
invalid address space.

With the implementation of para-virtualization and KVM-based virtualization,
Linux has support to represent the guest address space inside the kernel. This
would be one possiblity to get support for different address spaces in the
kernel. Alternatively, we can use the sparse memory model or HMM to create
``virtual'' memory nodes that correspond to an intermediate address space.

\myparagraph{Tracking Mappings}
\system uses designated mapping capabilities to track mappings. For every 
mapped object, there is a corresponding mapping capability, which is a 
descendant thereof. Therefore, the capability system is able to locate 
and invalidate all mappings when access to an object is revoked.
Note, translation structures effectively define an address space, and hence 
there is no difference between mappings of multi-level page tables, or 
actual frames.

Similar to the mapping capabilities, Linux uses the \verb|rmap| data structure
to store where a page of memory is mapped. This is already maintained for the
page cache, as well as guest memory pages. We can use this mechanisms to track
all mappings of a page in Linux.

\subsection{Model Population}
\label{sec:impl:modelpopulation}

The last part of the implementation describes how the model state is populated  
(\circled{3} in \autoref{fig:impl:overview}). There are two major sources of 
memory topology information building up the runtime representation: \ei static 
description of platforms (or parts there of), and \eii discovery mechanisms such 
as PCI or ACPI, which may instantiate predefined descriptions. 

\myparagraph{Static Platform Descriptions}
The memory topology of parts of the system -- or in the case of SoC the entire 
system -- is fixed and known in advance: for instance, the Xeon Phi 
co-processor has a defined number of cores and memory. We can therefore write 
down a description of the memory subsystem. For this, we use a domain 
specific language (DSL), which follows closely the syntax of the formal 
model, allows writing down the memory topology of the entire system, or its 
sub-components. The DSL compiler then produces a set of Prolog rules, which 
populate the model at runtime, either fully or in response to hardware 
discovery events. On Linux, we can use \verb|procfs| and \verb|sysfs|, as well as
device trees to obtain system topology descriptions.

\myparagraph{Using Static Descriptions: Code Generation}
From the static descriptions, we can pre-compute and enumerate the address 
spaces of the hardware component, or in the case of SoC platforms, the entire 
memory topology. The DSL compiler generates a set of data structures and 
code used by the reference monitor to instantiate the initial set of 
capabilities, verify address space compatibility in capability operations, 
translation tables, or functions to convert the canonical names into valid, 
local physical or virtual addresses. We evaluate this scenario 
in~\autoref{sec:eval:simulators}.

\myparagraph{Using Static Descriptions: Hardware Discovery}
In general, the configuration of a platform is known after device discovery 
mechanisms such as ACPI or PCI (if percent). During this process, the model is 
dynamically populated with the partial descriptions of its components: e.g. the 
ACPI table indicates the presence and version of an IOMMU, and in response the 
partial description of the IOMMU is instantiated and added to the model at 
runtime. A driver may update the model with more precise information, e.g. only 
the Xeon Phi driver knows the precise number of cores and memory size of the 
PCI Express attached co-processor.

\section{Evaluation}
\label{sec:evaluation}

In this section, we present a quantitative and qualitative performance 
evaluation of the address space and least-privilege authority model in \system.
The goal of this section is to establish the following:

\begin{myenumerate}
  \item The mechanism implementation results in a performant memory system 
  (\autoref{sec:eval:vmops}, 
  \autoref{sec:eval:appel}).
  \item The policy implementation produces usable results within 
  reasonable overheads (\autoref{sec:eval:realhw}).
    \item Qualitatively demonstrate, that the resulting system is able to handle 
  complex memory topologies (\autoref{sec:eval:simulators}).
  \end{myenumerate}

\myparagraph{Evaluation Platform}
All performance measurements are performed on a dual-socket server consisting of
two Intel Xeon E5-2670 v2 processors (\emph{Ivy-Bridge} micro-architecture) with
10 cores each. The machine has 256 GiB of main memory split equally into two
NUMA nodes. The machine runs in  ``\emph{performance mode}'', with
\emph{disabled} simultaneous multi-threading (SMT), Intel TurboBoost technology,
and Intel Speed Stepping, to ensure consistent measurements. The machine further
contains two Intel Xeon Phi co-processor 31S1 attached as a PCI Express 3.0
device. The co-processors have 57 cores with four hardware threads per core, and
8 GiB GDDR memory. The Intel VT-d~\cite{man:intel:vtd} (IOMMU) is enabled. We
use a vanilla Ubuntu 18.04 LTS with Linux kernel 4.15. For a fair comparison we
disable specter/meltdown mitigation as they slow down memory operations
significantly and Barrelfish doesn't implement them. Barrelfish and \system are
compiled in release mode.

\subsection{VM Ops - Map/Protect/Unmap}
\label{sec:eval:vmops}
In this part of the evaluation, we quantitatively evaluate the performance of 
\system's virtual memory operations in comparison to vanilla Barrelfish and 
Linux.

\myparagraph{Benchmark Methodology}
We compare the performance of the virtual memory operations \toperation{map},
\toperation{protect} and \toperation{unmap} for buffer sizes from 4 KiB to 64
GiB using one of the three native supported page sizes (4 KiB, 2 MiB and 1
GiB). On \system and Barrelfish, we use the default user-level virtual memory
management library, and on Linux we take the fastest of the measured different
techniques to map memory using anonymous memory (\toperation{mmap}), shared
memory objects (\toperation{shmfd}) or shared memory segment
(\toperation{shmat}). We exclude the allocation and clearing of backing memory
in this benchmark as it affects all systems the same and would dominate the
execution times.

\begin{figure}

 \begin{subfigure}[b]{\columnwidth}
    \includegraphics[height=2cm]{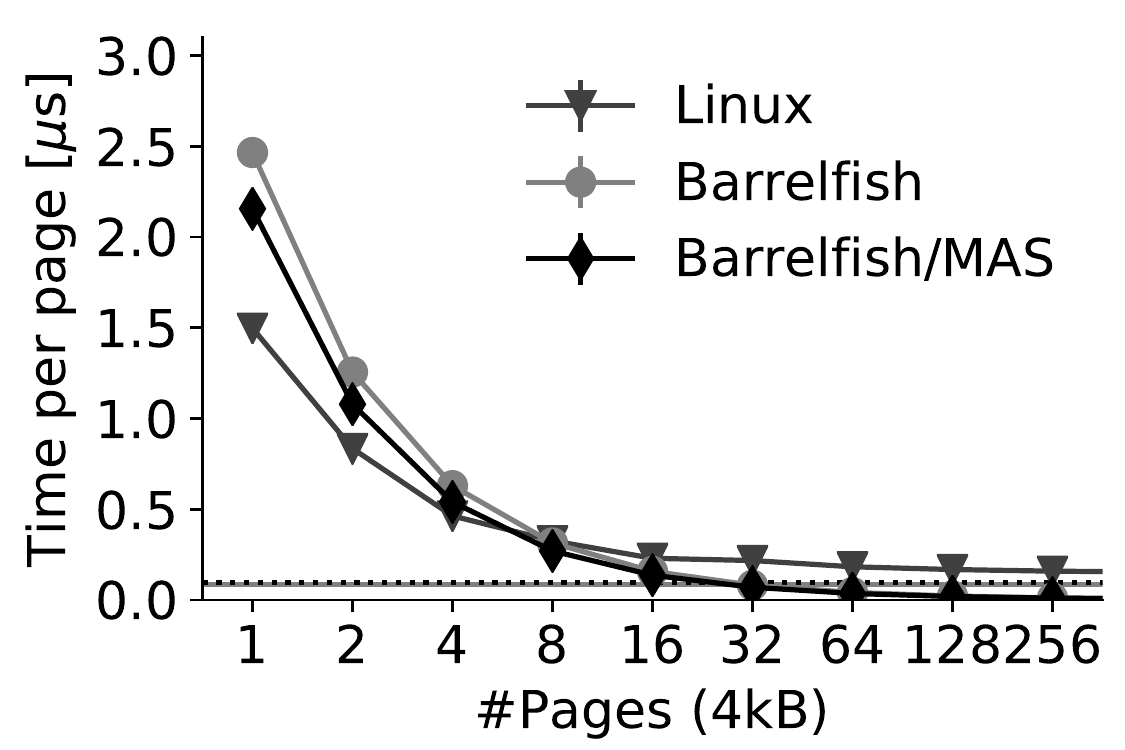}
    \includegraphics[height=2cm]{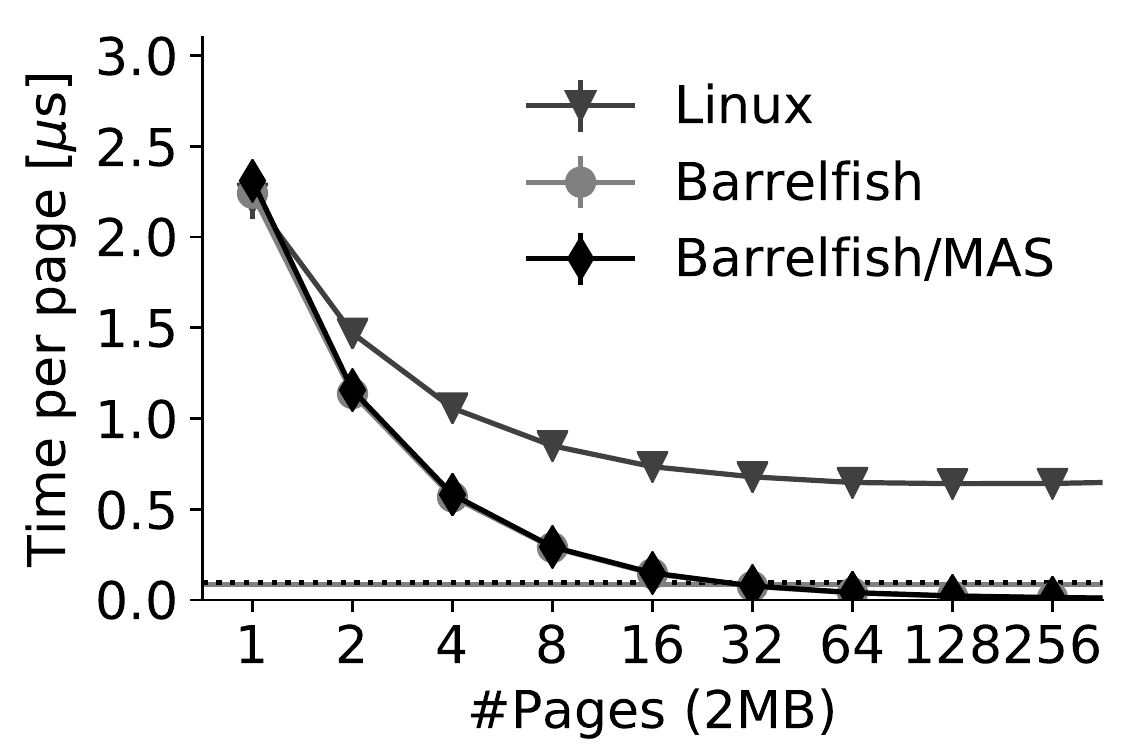}
    \includegraphics[height=2cm]{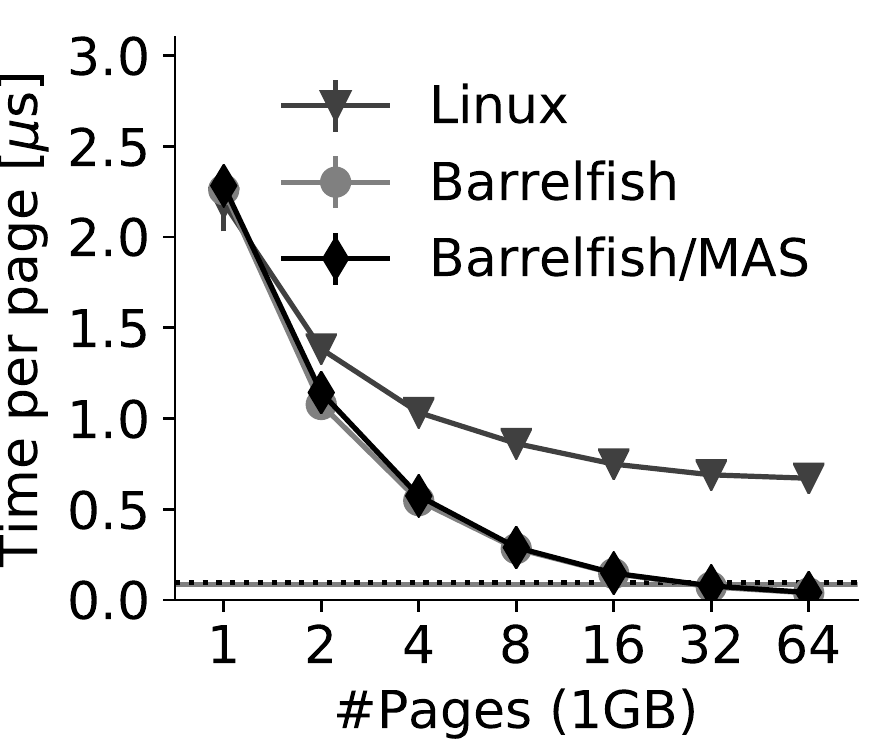}
    \vspace{-6mm}
    \caption{\toperation{map} Operation}
  \end{subfigure}

  \begin{subfigure}[b]{\columnwidth}
    \includegraphics[height=2cm]{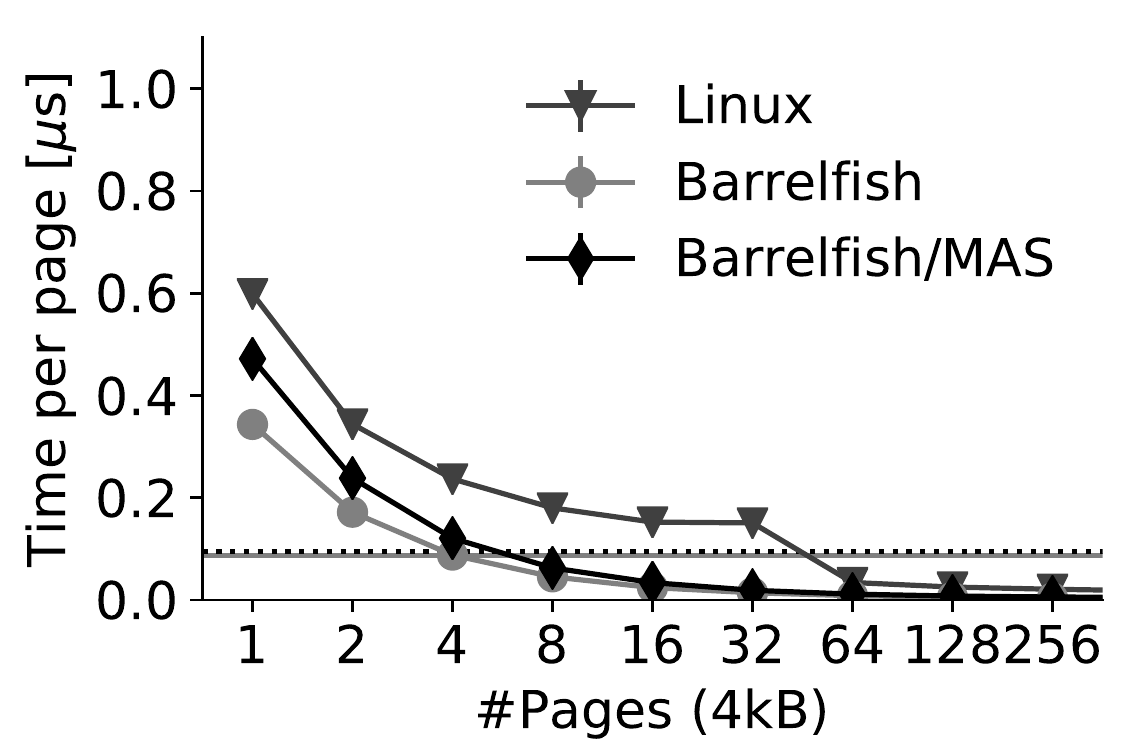}
    \includegraphics[height=2cm]{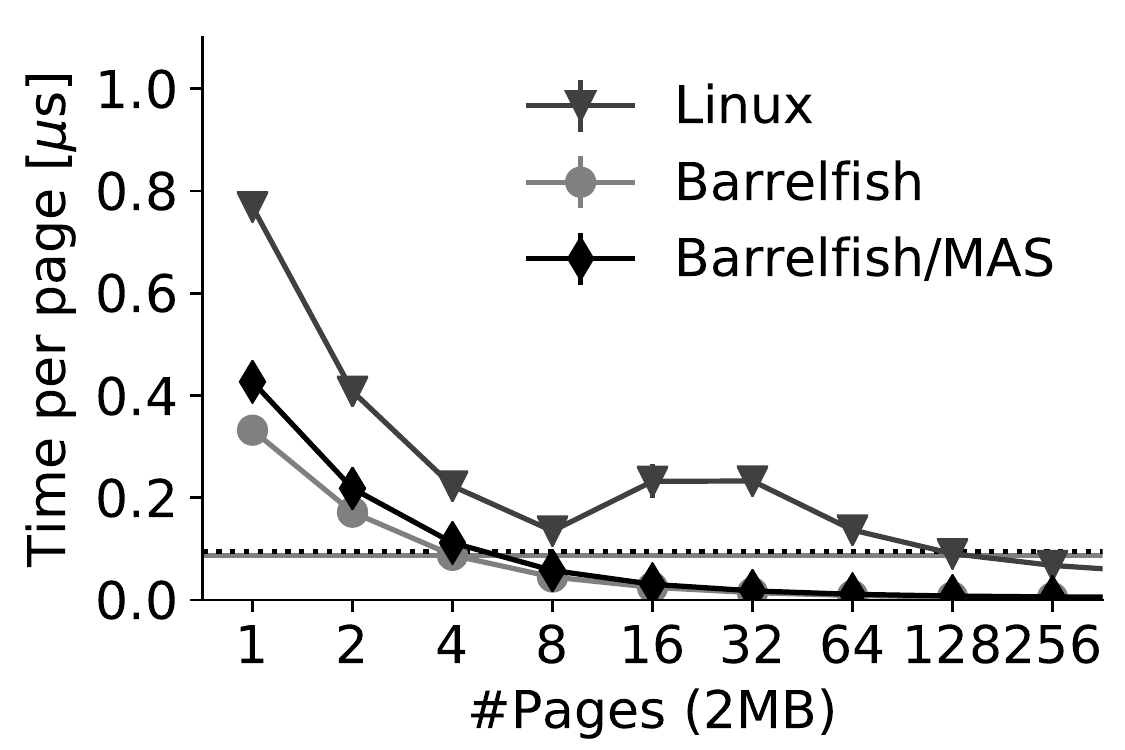}
    \includegraphics[height=2cm]{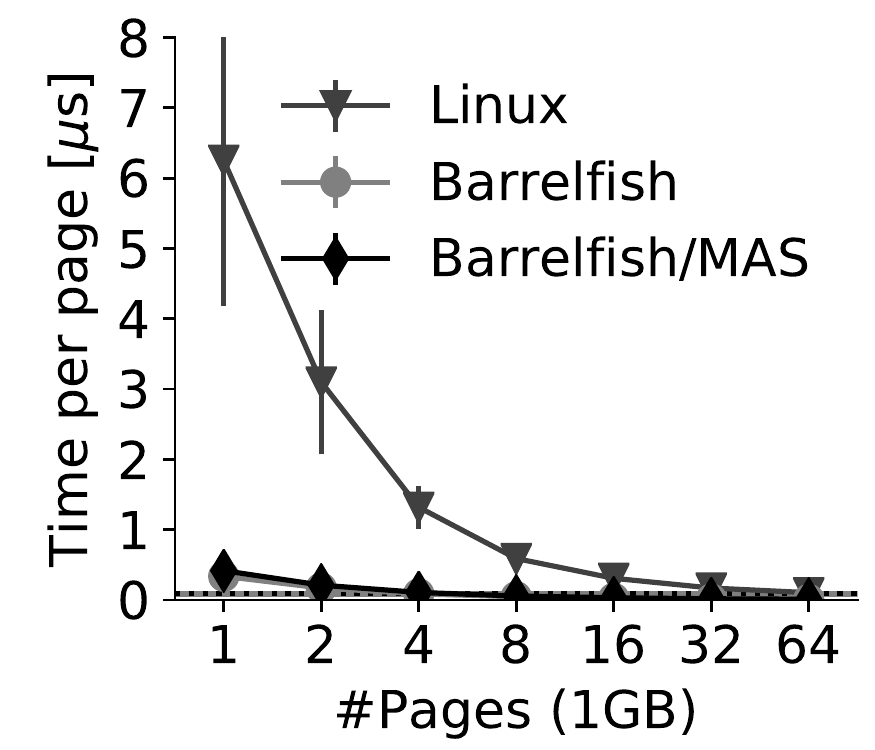}
    \vspace{-6mm}
    \caption{\toperation{protect} Operation}
  \end{subfigure}    
  \begin{subfigure}[b]{\columnwidth}
    \includegraphics[height=2cm]{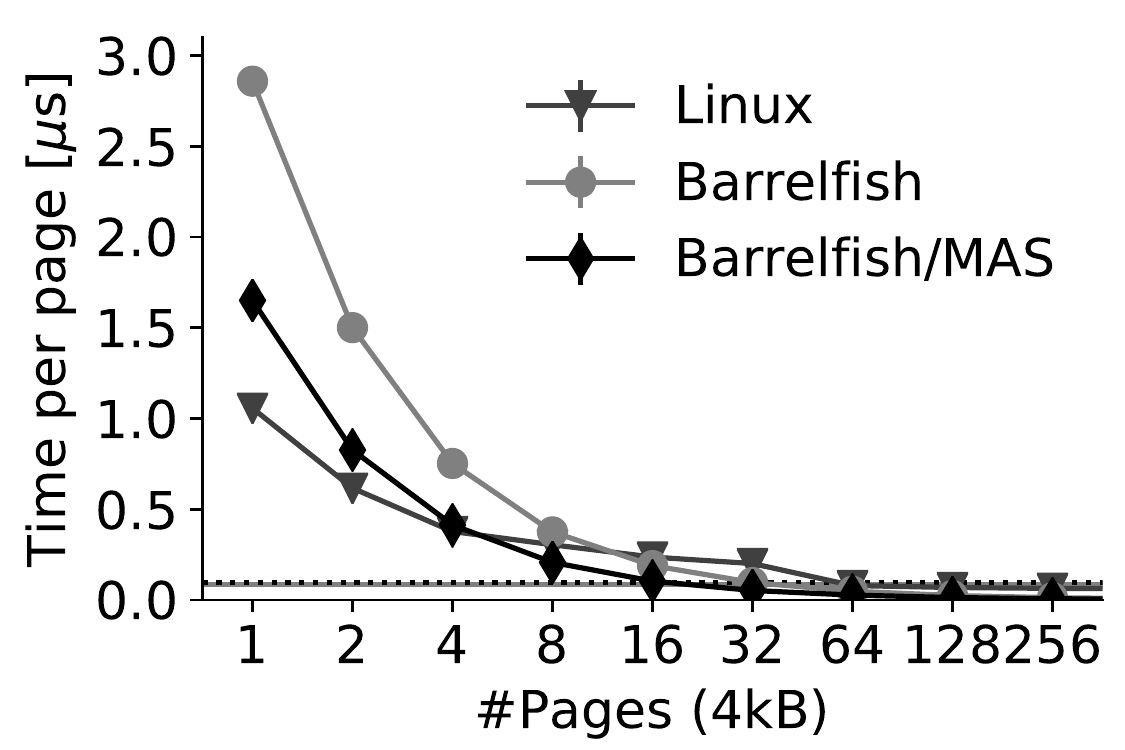}
    \includegraphics[height=2cm]{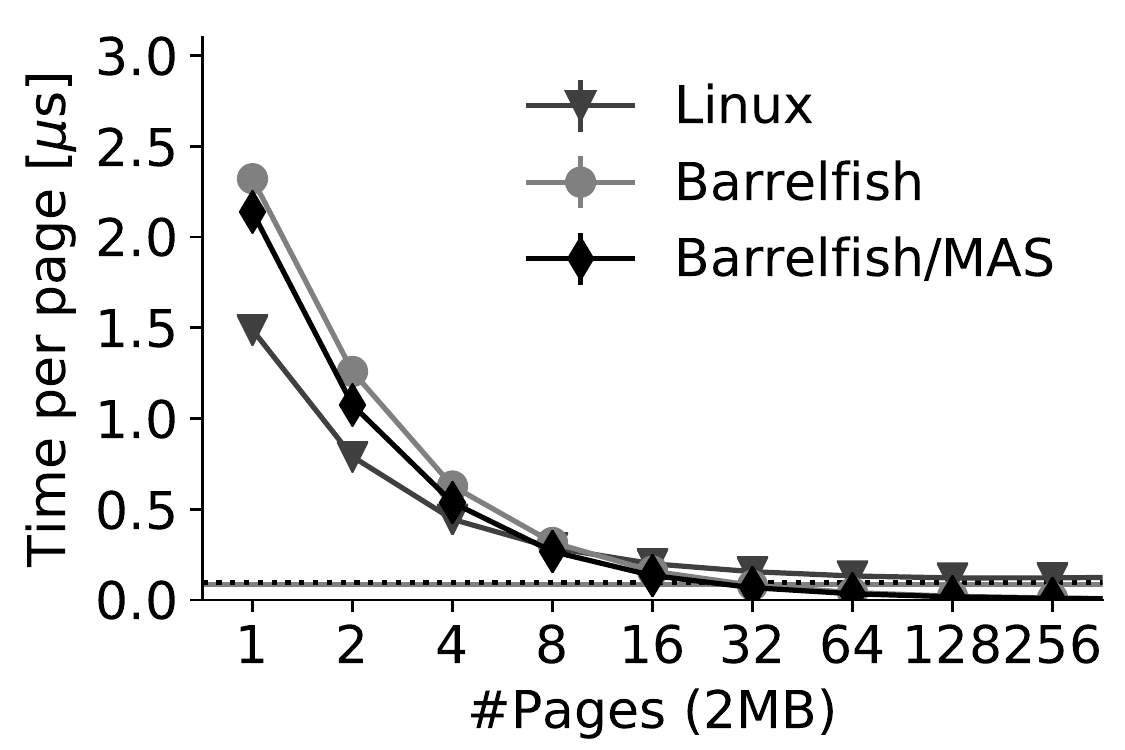}
    \includegraphics[height=2cm]{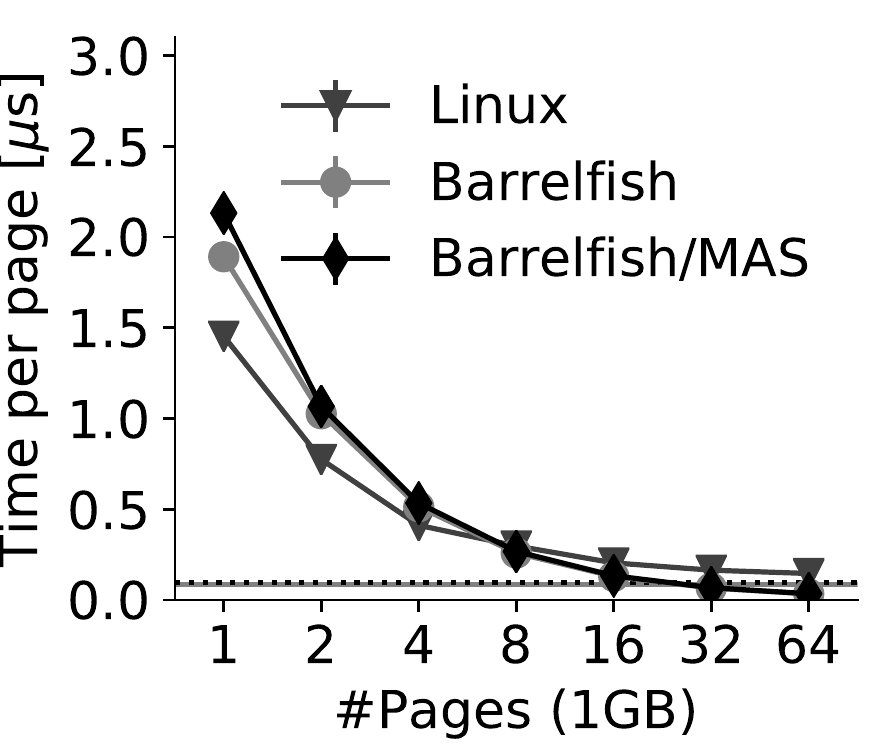}
    \vspace{-6mm}
    \caption{\toperation{unmap} Operation}
  \end{subfigure}
  \vspace{-4mm}
  \caption{Measured Latency per Page for the VM Operations on Linux, Barrelfish and 
  \system.}
  \label{fig:eval:memops}
\end{figure}

\begin{table}
  \begin{center}
    \begin{footnotesize}
      \begin{tabularx}{0.95\columnwidth}{llll}
        &  \toperation{map} & \toperation{protect} & \toperation{unmap} \\
        \hline
        4 KiB page       & Linux-shmfd & Linux-shmfd & Linux-shmat\\
        2 MiB large page & Linux-shmat & Linux-mmap  & Linux-shmat\\
        1 GiB huge page  & Linux-shmat & Linux-shmat & Linux-shmat\\
        \hline
      \end{tabularx}
    \end{footnotesize}
  \end{center}
  \vspace{-4mm}
  \caption{The Best Configuration of the Linux VM Operations.}
  \label{tab:impl:eval:linuxbest}
\end{table}

\myparagraph{Results}
\autoref{fig:eval:memops} contains the results of this evaluation for the three 
operations and page sizes. The graphs show the median latency (lower is better) 
and standard error per modified page table entry. We scale the number of changed 
page table entries. For Linux, we select the \emph{best} configuration as 
indicated in \autoref{tab:impl:eval:linuxbest}. We make the 
following observations:

\begin{myitemize}
  \item \emph{Amortization:\ } The general pattern is similar: the cost per page
  decreases with increasing numbers of affected pages. The cost of the virtual
  region management, syscall overhead, locating the page table entry is
  amortized among multiple pages, whose mappings are likely to be in consecutive
  page table entries.
    \item \toperation{map}.~Both, Barrelfish and \system have matching performance
  patterns, independent of the used page size. Linux is faster for mapping up to
  two 4 KiB pages. For larger pages Barrelfish (as well as \system) outperforms
  Linux. This is not an effect of our implementation but due to Linux allocating
  lower-level page tables, in case the super-page mapping needs to be broken up.
  Therefore, Linux allocates and clears memory to hold the page table. Zeroing a
  page can add up to 0.71$\mu s$ which is the difference we see in the graph.
  Both, Barrelfish and \system only have to create a new mapping capability and
  insert it into the MDB.

  \item \toperation{protect}.~ We observe very predictable patterns for
  Barrelfish and \system, where vanilla Barrelfish is slightly faster due to
  storing an explicit pointer to the page table directly in the mapping
  capability, whereas \system stores the canonical name which requires an
  address translation causing more work. In both cases, the mapping capability
  contains all information to perform the operation. Linux needs to walk the
  page table to locate the page table entry to be protected. This is again not
  an effect of the MAS extension but a difference between Linux and vanilla
  Barrelfish.

  \item \toperation{unmap}.~ Up to eight affected pages, Linux is faster than
  Barrelfish and \system, which both need to remove and delete the mapping
  capability from the MDB, which results in another syscall on Barrelfish 
  (\system removes this when clearing the page table entry). Removing the 
  mapping capability gets amortized when more pages are affected.  
  
\end{myitemize}

\myparagraph{Discussion}
\NH{    Under 6.1 Discussion. we have "can be implemented with fine 
granularity". Fine granularity is never mentioned before, it's not totally clear 
what is meant with "fine" in my opinion.}
In direct comparison with Barrelfish, we observe that \system is able to match
the performance in all cases. Moreover, the comparison with Linux shows, that
\system has comparable performance to a mainstream OS. We conclude that our
least-privilege access control model with support for multiple address spaces
can be implemented with fine granularity while maintaining competitive memory
management performance.

\subsection{VM Ops - Appel-Li Benchmark}
\label{sec:eval:appel}

The Appel-Li benchmark~\cite{Appel:1991:VMP} exercises the virtual memory 
subsystem with operations, which are relevant to tasks such as garbage 
collection or tracking page modifications.

\myparagraph{Benchmark Methodology}
The benchmark consists of the following three experiments:
\begin{myenumerate}
  \item \emph{prot1-trap-unprot.\ } 
  Randomly pick a page of memory, write-protect the page, write to it, take 
  a trap, unprotect the page, continue with next page.
    \item \emph{protN-trap-unprot.\ } 
  Write-protect 512 pages of memory at once, write to each page of memory in
  turn, taking a trap and unprotecting the page.
  
  \item \emph{trap only.\ } 
  Pick a protected page, write to it and take the trap continue with next 
  page without changing any permissions.
\end{myenumerate}
We run this benchmark on Barrelfish and \system. In addition, we compare to
Linux as a frame of reference. On Barrelfish and \system the numbers include 
the cost of virtual address space accounting in userspace.

\begin{figure}
  \begin{center}
    \includegraphics[width=\columnwidth]{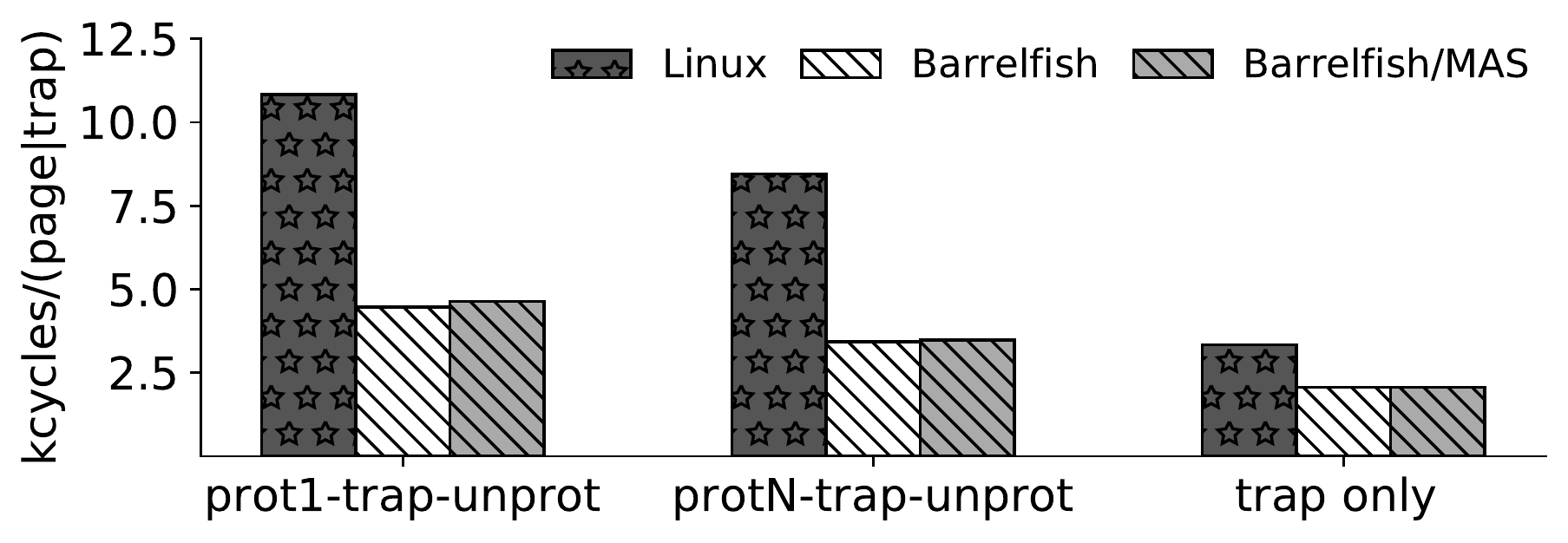}
  \end{center}
  \vspace{-4mm}
  \caption{Appel-Li Benchmark on \system and Linux.}
  \label{fig:eval:appel-li}
\end{figure}

\myparagraph{Results}
We show the benchmark results in \autoref{fig:eval:appel-li}. Each bar 
corresponds to a different OS and represents the time taken per 
page. The three bar groups represent the three benchmark experiments. 
The standard error is less than 0.5\%. We make the following observations:

\begin{myitemize}
  \item \emph{Barrelfish vs. \system.~} Direct comparison shows a slowdown of 
  less than 5\% for \system vs. Barrelfish. The trap performance of both systems 
  is the same.
    \item \emph{Linux vs Barrelfish.~} Barrelfish outperforms Linux in all
  experiments. Barrelfish can use its capability system to efficiently find
  the page table that has to be modified while Linux needs to walk the page
  table tree. Furthermore Barrelfish reflects the trap directly to user-space
  without checking whether the faulting address has been previously
  allocated~\cite{Bar:2000:MEA}. This applies to \system as well as vanilla
  Barrelfish and is independent of our extension.
    \item \emph{Batching.~} The protection of 512 pages in one syscall 
  (\emph{protN-trap-unprot}) amortizes the total syscall overheads, which
  reduce the time per page on all systems by 600-2000 cycles.
\end{myitemize}

\myparagraph{Discussion}
In this evaluation, we show that \system is able to match the performance of
Barrelfish with a maximum overhead of less than 5\%, despite support for
explicit address spaces. The comparison to Linux again shows that \system's
memory operation performance is competitive to that of a mainstream OS.

\subsection{Dynamic Updates of Translation Tables}
\label{sec:eval:realhw}

In this evaluation, we investigate the overheads of the model runtime 
representation and the translation unit re-configuration following the principle 
of least-privilege.

\myparagraph{Benchmark Methodology}
This benchmark models an offload-scenario, where an application workload wants 
to make use of a co-processor attached to PCI Express. We use the Xeon Phi 
co-processor for this purpose. We are interested in the sequence of 
initialization steps to establish a shared buffer between the CPU cores and the 
co-processor:

\begin{myenumerate}
  \item \emph{Model Query.~} Evaluate the runtime representation to find a 
  suitable memory region and needed re-configuration steps. 
    \item \emph{Allocate and Map.~} Request memory from the allocator 
  and map it into the application's virtual address space.
    \item \emph{Program Translation Units.~} Re-configure the translation 
  units indicated in the model query response. Here, this includes 
  \ei the IOMMU, and \eii the SMPT of the co-processor.
\end{myenumerate}

We profile the execution of these steps and measure the time it takes to perform
each step individually. We evaluate two mechanisms to program the IOMMU, \ei to
use capability invocations directly, and \eii use an RPC to the IOMMU service
acting as a reference monitor. The buffer size used is 8 MiB. As a frame of
reference, measure the time it takes to just allocate and map memory on both
Linux (using \toperation{mmap}) and vanilla Barrelfish.

\begin{figure}
  \includegraphics[width=\columnwidth]{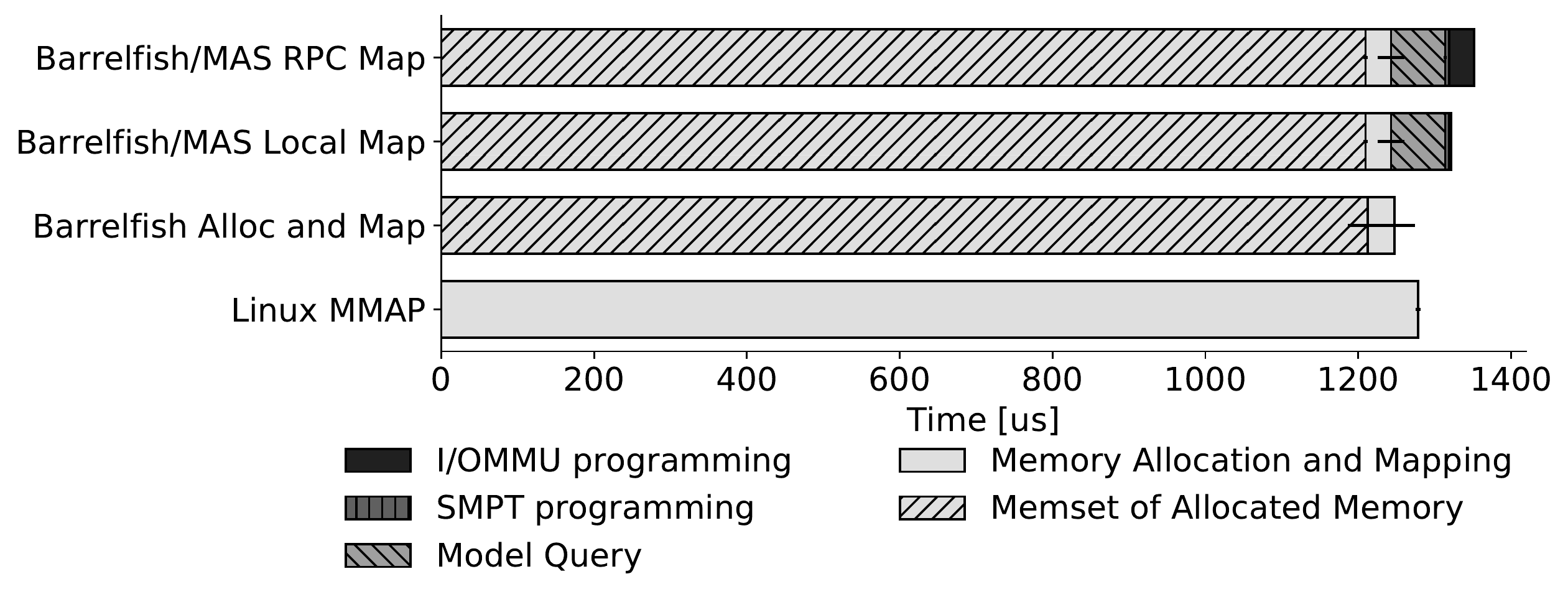}
  \vspace{-5mm}
  \caption{Breakdown of the Offloading Scenario.}
  \label{fig:impl:eval:bufalloc}
\end{figure}

\myparagraph{Results}
The breakdown of the operation into the steps is shown in
\autoref{fig:impl:eval:bufalloc}. We show both the numbers for both mechanisms
to program the IOMMU, and for comparison, we include the time it takes to just
allocate and map the memory on vanilla Barrelfish and Linux. The x-axis
represents the measured times in $\mu s$. We make the following observations:

\begin{myitemize}
  \item \emph{Memory Allocation and Mapping.~} All three OSes use about the same 
  time to allocate and map the required memory region, which accounts for the 
  majority of the profiled time. It is dominated by zeroing the newly 
  allocated memory.
  
  \item \emph{Model Query.~} Evaluating the model at runtime accounts for less 
  than 5\% of the total runtime. 
  
  \item \emph{SMPT Configuration.~} Programming the SMPT of the co-processor 
  uses less than 0.3\% of the runtime. 
 
  \item \emph{IOMMU Programming.~} The configuration of the IOMMU using direct 
  capability invocations is fast (0.2\% of the runtime). When using the RPC to 
  the IOMMU reference monitor, this requires capability transfers which 
  corresponds to about 3\% of the execution time. 
\end{myitemize}

Overall, the resulting overhead for the model query and the address space 
configuration accounts for $5.7\%$. There is no significant difference 
in the memory allocation and mapping times of \system compared to Barrelfish 
and Linux.

\myparagraph{Discussion}
In this evaluation, we have shown that it is possible to efficiently implement 
a representation of our executable model in an operating system and reconfigure 
address spaces following the principle of least-privilege. Moreover, subsequent 
allocations may use the cached results of the model query, reducing the overhead 
even further. Note, that the query merely indicate the operations to be carried 
out, but the capability system enforces the integrity thereof.

\subsection{Correctness on Simulated Platforms}
\label{sec:eval:simulators}

In this evaluation, we qualitatively show the application and integration 
of the address space model into the OS toolchain to \emph{generate} low-level, 
platform-specific OS code and data structures. By doing that we show, that our
implementation is functional even when run on simulated platforms with unusual 
address space topologies not supported by other systems. While these simulated 
platforms are extreme, they include other real systems such as those 
with secure co-processors.

\begin{figure}[!t]
  \includegraphics{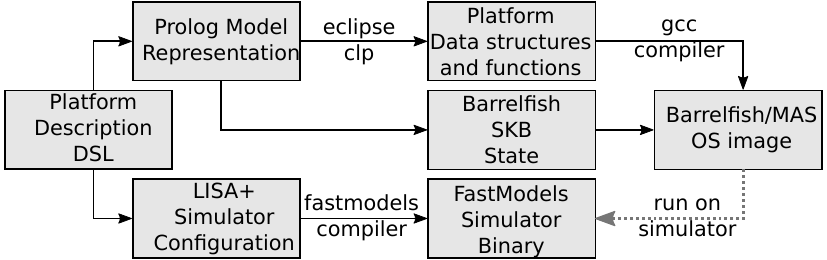}
  \caption{Running \system on an ARM FastModels~\cite{fastmodels} Platform Based 
    on a Hardware Description. }
  \label{fig:impl:eval:simulator}
\end{figure}

\myparagraph{Evaluation Methodology}
We design and build the toolchain illustrated in 
\autoref{fig:impl:eval:simulator} and write a series of different platform 
descriptions using a DSL. These platform descriptions then specify the memory 
topology of the simulated platforms. The DSL compiler then generates:

\begin{myenumerate}
  \item \emph{Executable Model.~} A runtime representation of the memory 
  topology model, and
  \item \emph{Simulator Configuration.~} The LISA+ hardware description 
  that configures the ARM FastModels simulator~\cite{fastmodels}.
\end{myenumerate}

The generated runtime representation of the topology model then acts as the 
initial state for the Barrelfish SKB, and is used to generate low-level OS code 
and data structures, which are compiled and linked into a platform-specific 
\system OS image.

We mention four example configurations we tested for this evaluation. 
\autoref{fig:impl:eval:simulatorconfig} shows an 
illustration of the simulated platform, which consists of two ARM Cortex A57 
processors, each having a 
configurable local memory map which defines at which addresses they see the DRAM 
regions (and the rest of the system in general) in their local address space. We 
evaluated the following configurations:

\begin{figure}[!t]
  \includegraphics{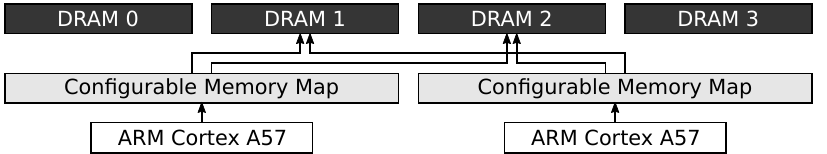}
  \caption{FastModels Simulator Configuration}
  \label{fig:impl:eval:simulatorconfig}
\end{figure}

\begin{myenumerate}
  \item \emph{Uniform} Both cores have an identical memory map.
  \item \emph{Swapped} DRAM is split in two halves, where each core sees the two 
  halves at swapped address ranges. 
  \item \emph{Private} One shared memory region, and each core further has 
  a private memory region, inaccessible by the other. 
  \item \emph{Private Swapped} Combines the \emph{swapped} and \emph{private} 
  setups: shared memory with swapped views, and private memory per core.
\end{myenumerate}

\myparagraph{Results}
During out experiments, we managed to compile \system and run it successfully on 
all tested platform configurations. This includes various memory management 
tasks and shared-memory message passing between the cores. There was no 
programmer effort required, besides writing the platform description.

\myparagraph{Discussion}
We know of no other current OS designs which can manage
memory globally in all these cases. Popcorn Linux~\cite{Barbalace:2015:PBP} and
Barrelfish have limited support for case 3; while regular Linux
and seL4 only support case 1. In contrast, \system supports all four cases.

\system is able to boot and manage memory on all platforms without 
modifications, regardless of the topology.

\subsection{Evaluation Summary}
\label{sec:impl:conclusion}

In this evaluation, we have shown that it is possible to efficiently
implement the address space model and least-privilege memory management 
in an OS. We have quantitatively evaluated \system's virtual memory system, the 
reconfiguration operations, and analyzed the space and runtime complexity of 
maintaining kernel state. 

Moreover, we have seen that \system is able to handle complex and non-standard 
memory topologies by strictly using the memory object's 
canonical name in the capability system, and generated translation functions 
which further convert this canonical name to a valid local address

\section{Conclusion}
\label{sec:conclusion}

In this paper, we made the case to bring back the concept of a reference monitor
to mediate access to memory resource on modern, heterogeneous platforms. We presented
a fine-grained, realistic memory protection model based on which we can extend the 
reference monitor to include all memory translation and protection hardware present 
in the system. This allows systems software to adapt their access control model and 
catch up with the complexity of modern hardware.

We have shown that our design is applicable to any OS, regardless of its 
architecture. We have developed an executable specification of a reference
monitor including the state, operations and authority, on which we have based our 
prototype implementation in \system. Not only can this memory protection model 
eliminate three different classes of bugs and vulnerabilities, but there is also no 
inherent performance overhead in implementing it in an operating system. Moreover, 
based on trusted hardware specifications we can increase the level of automation and 
generate low-level operating systems code. We believe that our approach can lay the 
foundation for both fully verified systems and more reliable memory management in 
existing systems.

We plan to open-source the reference monitor and \system implementations.

\bibliographystyle{plain}
\bibliography{references}

\begin{thebibliography}{10}

\bibitem{Achermann:2017:FMA}
Reto Achermann, Lukas Humbel, David Cock, and Timothy Roscoe.
\newblock {Formalizing Memory Accesses and Interrupts}.
\newblock In {\em Proceedings of the 2nd Workshop on Models for Formal Analysis
  of Real Systems}, MARS 2017, pages 66--116, 2017.

\bibitem{Achermann:2018:PAR}
Reto Achermann, Lukas Humbel, David Cock, and Timothy Roscoe.
\newblock {Physical Addressing on Real Hardware in Isabelle/HOL}.
\newblock In {\em Proceedings of the 9th International Conference on
  Interactive Theorem Proving}, ITP'18, pages 1--19, Oxford, United Kingdom,
  2018. Springer International Publishing.

\bibitem{Alam:2017:DVM}
Hanna Alam, Tianhao Zhang, Mattan Erez, and Yoav Etsion.
\newblock {Do-It-Yourself Virtual Memory Translation}.
\newblock In {\em Proceedings of the 44th Annual International Symposium on
  Computer Architecture}, ISCA '17, pages 457--468, New York, NY, USA, 2017.
  ACM.

\bibitem{Anderson:1972:reference_monitor}
James~P. Anderson.
\newblock {Computer Security Technology Planning Study}.
\newblock Technical Report ESD-TR-73-51, Vol. I, AD-758 206, Electronic Systems
  Division, Deputy for Command and Management Systems HQ Electronic Systems
  Division (AFSC), L. G. Hanscom Field, Bedford, Massachusetts 01730, USA, 10
  1972.

\bibitem{Appel:1991:VMP}
Andrew~W. Appel and Kai Li.
\newblock {Virtual Memory Primitives for User Programs}.
\newblock In {\em Proceedings of the Fourth International Conference on
  Architectural Support for Programming Languages and Operating Systems},
  ASPLOS IV, pages 96--107, New York, NY, USA, 1991. ACM.

\bibitem{man:arm:trustzone}
{ARM Ltd}.
\newblock {\em {ARM Security Technology - Building a Secure System using
  TrustZone Technology}}, prd29-genc-009492c edition, 4 2009.

\bibitem{fastmodels}
{ARM Ltd.}
\newblock {Development Tools and Software: Fast Models}.
\newblock
  \url{https://www.arm.com/products/development-tools/simulation/fast-models},
  8 2019.

\bibitem{Ausavarungnirun:2017:MGM}
Rachata Ausavarungnirun, Joshua Landgraf, Vance Miller, Saugata Ghose, Jayneel
  Gandhi, Christopher~J. Rossbach, and Onur Mutlu.
\newblock {Mosaic: A GPU Memory Manager with Application-transparent Support
  for Multiple Page Sizes}.
\newblock In {\em Proceedings of the 50th Annual IEEE/ACM International
  Symposium on Microarchitecture}, MICRO-50 '17, pages 136--150, New York, NY,
  USA, 2017. ACM.

\bibitem{Bar:2000:MEA}
Moshe Bar.
\newblock {The Linux Signals Handling Model}.
\newblock {\em Linux Journal}, 5 2000.
\newblock \url{https://www.linuxjournal.com/article/3985}.

\bibitem{Barbalace:2017:TTO}
Antonio Barbalace, Anthony Iliopoulos, Holm Rauchfuss, and Goetz Brasche.
\newblock {It's Time to Think About an Operating System for Near Data
  Processing Architectures}.
\newblock In {\em Proceedings of the 16th Workshop on Hot Topics in Operating
  Systems}, HotOS '17, pages 56--61, New York, NY, USA, 2017. ACM.

\bibitem{Barbalace:2015:PBP}
Antonio Barbalace, Marina Sadini, Saif Ansary, Christopher Jelesnianski, Akshay
  Ravichandran, Cagil Kendir, Alastair Murray, and Binoy Ravindran.
\newblock {Popcorn: Bridging the Programmability Gap in heterogeneous-ISA
  Platforms}.
\newblock In {\em Proceedings of the Tenth European Conference on Computer
  Systems}, EuroSys '15, pages 29:1--29:16, New York, NY, USA, 2015. ACM.

\bibitem{Baumann:2009:MNO}
Andrew Baumann, Paul Barham, Pierre-Evariste Dagand, Tim Harris, Rebecca
  Isaacs, Simon Peter, Timothy Roscoe, Adrian Sch\"{u}pbach, and Akhilesh
  Singhania.
\newblock {The Multikernel: A New OS Architecture for Scalable Multicore
  Systems}.
\newblock In {\em Proceedings of the ACM SIGOPS 22nd Symposium on Operating
  Systems Principles}, SOSP '09, pages 29--44, New York, NY, USA, 2009. ACM.

\bibitem{Bigs:2018:JMOS}
Simon Biggs, Damon Lee, and Gernot Heiser.
\newblock The jury is in: Monolithic os design is flawed: Microkernel-based
  designs improve security.
\newblock In {\em Proceedings of the 9th Asia-Pacific Workshop on Systems},
  APSys ’18, New York, NY, USA, 2018. Association for Computing Machinery.

\bibitem{Chester:2018:ECV}
Adam Chester.
\newblock {Exploiting CVE-2018-1038 - Total Meltdown}.
\newblock Online. \url{https://blog.xpnsec.com/total-meltdown-cve-2018-1038/},
  4 2018.

\bibitem{Cock:2008:SMS}
David Cock, Gerwin Klein, and Thomas Sewell.
\newblock {Secure Microkernels, State Monads and Scalable Refinement}.
\newblock In {\em Proceedings of the 21st International Conference on Theorem
  Proving in Higher Order Logics}, TPHOLs '08, pages 167--182, Berlin,
  Heidelberg, 2008. Springer-Verlag.

\bibitem{Dautenhahn:2015:NKO}
Nathan Dautenhahn, Theodoros Kasampalis, Will Dietz, John Criswell, and Vikram
  Adve.
\newblock Nested kernel: An operating system architecture for intra-kernel
  privilege separation.
\newblock In {\em Proceedings of the Twentieth International Conference on
  Architectural Support for Programming Languages and Operating Systems},
  ASPLOS '15, pages 191--206, New York, NY, USA, 2015. ACM.

\bibitem{Derrin:2006:RMA}
Philip Derrin, Kevin Elphinstone, Gerwin Klein, David Cock, and Manuel M.~T.
  Chakravarty.
\newblock {Running the Manual: An Approach to High-assurance Microkernel
  Development}.
\newblock In {\em Proceedings of the 2006 ACM SIGPLAN Workshop on Haskell},
  Haskell '06, pages 60--71, New York, NY, USA, 2006. ACM.

\bibitem{Elkaduwe:2008:VPM}
Dhammika Elkaduwe, Gerwin Klein, and Kevin Elphinstone.
\newblock Verified protection model of the sel4 microkernel.
\newblock In {\em Proceedings of the 2nd International Conference on Verified
  Software: Theories, Tools, Experiments}, VSTTE '08, pages 99--114, Berlin,
  Heidelberg, 2008. Springer-Verlag.

\bibitem{Ferraiuolo:2017:KUV}
Andrew Ferraiuolo, Andrew Baumann, Chris Hawblitzel, and Bryan Parno.
\newblock {Komodo: Using Verification to Disentangle Secure-enclave Hardware
  from Software}.
\newblock In {\em Proceedings of the 26th Symposium on Operating Systems
  Principles}, SOSP '17, pages 287--305, New York, NY, USA, 2017. ACM.

\bibitem{Gerber:2015:YPP}
Simon Gerber, Gerd Zellweger, Reto Achermann, Kornilios Kourtis, Timothy
  Roscoe, and Dejan Milojicic.
\newblock {Not Your Parents' Physical Address Space}.
\newblock In {\em Proceedings of the 15th USENIX Conference on Hot Topics in
  Operating Systems}, HOTOS'15, pages 16--16, Berkeley, CA, USA, 2015. USENIX
  Association.

\bibitem{Gong:2019:EQW}
Xiling Gong.
\newblock {Exploiting Qualcomm WLAN and Modem Over the Air}.
\newblock In {\em Proceedings of the BlackHat USA 2019}, 2019.

\bibitem{Gu:2016:CEA}
Ronghui Gu, Zhong Shao, Hao Chen, Xiongnan Wu, Jieung Kim, Vilhelm Sj\"{o}berg,
  and David Costanzo.
\newblock {CertiKOS: An Extensible Architecture for Building Certified
  Concurrent OS Kernels}.
\newblock In {\em Proceedings of the 12th USENIX Conference on Operating
  Systems Design and Implementation}, OSDI'16, pages 653--669, Berkeley, CA,
  USA, 2016. USENIX Association.

\bibitem{Hillenbrand:2017:MPM}
Marius Hillenbrand, Mathias Gottschlag, Jens Kehne, and Frank Bellosa.
\newblock {Multiple Physical Mappings: Dynamic DRAM Channel Sharing and
  Partitioning}.
\newblock In {\em Proceedings of the 8th Asia-Pacific Workshop on Systems},
  APSys '17, pages 21:1--21:9, Mumbai, India, 2017.

\bibitem{man:hsa}
{HSA Foundation}.
\newblock {\em {HSA Runtime Programmer's Reference Manual}}, version: 1.1.4
  edition, 10 2016.

\bibitem{Huang:2016:ESL}
Jian Huang, Moinuddin~K. Qureshi, and Karsten Schwan.
\newblock {An Evolutionary Study of Linux Memory Management for Fun and
  Profit}.
\newblock In {\em Proceedings of the 2016 USENIX Conference on Usenix Annual
  Technical Conference}, USENIX ATC '16, pages 465--478, Berkeley, CA, USA,
  2016. USENIX Association.

\bibitem{man:intel:vtd}
{Intel Corporation}.
\newblock {\em {Intel Virtualization Technology for Directed I/O - Architecture
  Specification}}, d51397-011, revision 3.1 edition, 6 2019.

\bibitem{man:opencl}
{Khronos OpenCL Working Group}.
\newblock {\em The OpenCL Specification}, version: 2.1, document revision: 24
  edition, 2 2018.

\bibitem{Klein:2009:SFV}
Gerwin Klein, Kevin Elphinstone, Gernot Heiser, June Andronick, David Cock,
  Philip Derrin, Dhammika Elkaduwe, Kai Engelhardt, Rafal Kolanski, Michael
  Norrish, Thomas Sewell, Harvey Tuch, and Simon Winwood.
\newblock {seL4: Formal Verification of an OS Kernel}.
\newblock In {\em Proceedings of the ACM SIGOPS 22nd Symposium on Operating
  Systems Principles}, SOSP '09, pages 207--220, New York, NY, USA, 2009. ACM.

\bibitem{Lampson:1974:Protection}
Butler~W Lampson.
\newblock Protection.
\newblock {\em {ACM SIGOPS Operating Systems Review}}, 8(1):18--24, 1974.

\bibitem{Lee:2014:VIL}
Janghaeng Lee, Mehrzad Samadi, and Scott Mahlke.
\newblock {VAST: The Illusion of a Large Memory Space for GPUs}.
\newblock In {\em Proceedings of the 23rd International Conference on Parallel
  Architectures and Compilation}, PACT '14, pages 443--454, New York, NY, USA,
  2014. ACM.

\bibitem{man:linux:hmm}
{Linux Kernel Documentation}.
\newblock {\em {Heterogeneous Memory Management (HMM)}}, version 5.0 edition, 4
  2019.

\bibitem{Markettos:2019:TEV}
A~Theodore Markettos, Colin Rothwell, Brett~F Gutstein, Allison Pearce, Peter~G
  Neumann, Simon~W Moore, and Robert~NM Watson.
\newblock {Thunderclap: Exploring Vulnerabilities in Operating System IOMMU
  Protection via DMA from Untrustworthy Peripherals}.
\newblock In {\em NDSS}, 2019.

\bibitem{Markuze:2016:TIP}
Alex Markuze, Adam Morrison, and Dan Tsafrir.
\newblock {True IOMMU Protection from DMA Attacks: When Copy is Faster Than
  Zero Copy}.
\newblock In {\em Proceedings of the Twenty-First International Conference on
  Architectural Support for Programming Languages and Operating Systems},
  ASPLOS '16, pages 249--262, New York, NY, USA, 2016. ACM.

\bibitem{Morgan:2016:BIP}
Benot Morgan, Eric Alata, Vincent Nicomette, and Mohamed Kaaniche.
\newblock {Bypassing IOMMU Protection against I/O Attacks}.
\newblock In {\em 2016 Seventh Latin-American Symposium on Dependable Computing
  (LADC)}, pages 145--150, 10 2016.

\bibitem{Morgan:2018:IPIO}
Benot Morgan, Eric Alata, Vincent Nicomette, and Mohamed Kaaniche.
\newblock {IOMMU Protection Against I/O Attacks: A Vulnerability and a Proof of
  Concept}.
\newblock {\em Journal of the Brazilian Computer Society}, 24(1):2, 1 2018.

\bibitem{CVE-2011-1898}
NATIONAL VULNERABILITY~DATABASE NVD.
\newblock {CVE-2011-1898}.
\newblock Online, 8 2011.

\bibitem{CVE-2013-4329}
NATIONAL VULNERABILITY~DATABASE NVD.
\newblock {CVE-2013-4329}.
\newblock Online, 9 2013.

\bibitem{CVE-2014-0972}
NATIONAL VULNERABILITY~DATABASE NVD.
\newblock {CVE-2014-0972}.
\newblock Online, 8 2014.

\bibitem{CVE-2014-3601}
NATIONAL VULNERABILITY~DATABASE NVD.
\newblock {CVE-2014-3601}.
\newblock Online, 8 2014.

\bibitem{CVE-2014-9888}
NATIONAL VULNERABILITY~DATABASE NVD.
\newblock {CVE-2014-9888}.
\newblock Online, 8 2014.

\bibitem{CVE-2015-6994}
NATIONAL VULNERABILITY~DATABASE NVD.
\newblock {CVE-2015-6994}.
\newblock Online, 1 2017.

\bibitem{CVE-2016-5349}
NATIONAL VULNERABILITY~DATABASE NVD.
\newblock {CVE-2016-5349}.
\newblock Online, 4 2017.

\bibitem{CVE-2017-12188}
NATIONAL VULNERABILITY~DATABASE NVD.
\newblock {CVE-2017-12188}.
\newblock Online, 10 2017.

\bibitem{CVE-2018-1038}
NATIONAL VULNERABILITY~DATABASE NVD.
\newblock {CVE-2018-1038}.
\newblock Online, 8 2018.

\bibitem{CVE-2015-4421}
NATIONAL VULNERABILITY~DATABASE NVD.
\newblock {CVE-2015-4421}.
\newblock Online, 5 2019.

\bibitem{CVE-2015-4422}
NATIONAL VULNERABILITY~DATABASE NVD.
\newblock {CVE-2015-4422}.
\newblock Online, 5 2019.

\bibitem{CVE-2019-10538}
NATIONAL VULNERABILITY~DATABASE NVD.
\newblock {CVE-2019-10538 - Modem into Linux Kernel issue}.
\newblock Online, 8 2019.

\bibitem{CVE-2019-10539}
NATIONAL VULNERABILITY~DATABASE NVD.
\newblock {CVE-2019-10539 - Compromise WLAN Issue}.
\newblock Online, 8 2019.

\bibitem{CVE-2019-10540}
NATIONAL VULNERABILITY~DATABASE NVD.
\newblock {CVE-2019-10540 - WLAN into Modem issue}.
\newblock Online, 8 2019.

\bibitem{man:nvidia:UMCUDA}
{NVIDIA Corporation}.
\newblock {\em {Unified Memory in CUDA 6}}, 11 2013.

\bibitem{Patterson:1997:CIR}
David Patterson, Thomas Anderson, Neal Cardwell, Richard Fromm, Kimberly
  Keeton, Christoforos Kozyrakis, Randi Thomas, and Katherine Yelick.
\newblock {A Case for Intelligent RAM}.
\newblock {\em IEEE Micro}, 17(2):34--44, 3 1997.

\bibitem{Romanescu:2010:SDV}
Bogdan~F. Romanescu, Alvin~R. Lebeck, and Daniel~J. Sorin.
\newblock {Specifying and Dynamically Verifying Address Translation-aware
  Memory Consistency}.
\newblock In {\em Proceedings of the Fifteenth Edition of ASPLOS on
  Architectural Support for Programming Languages and Operating Systems},
  ASPLOS XV, pages 323--334, New York, NY, USA, 2010. ACM.

\bibitem{Schnarz:2014:TAR}
Pierre Schnarz, Joachim Wietzke, and Ingo Stengel.
\newblock Towards attacks on restricted memory areas through co-processors in
  embedded multi-os environments via malicious firmware injection.
\newblock In {\em Proceedings of the First Workshop on Cryptography and
  Security in Computing Systems}, CS2 '14, pages 25--30, New York, NY, USA,
  2014. ACM.

\bibitem{Schupbach:2011:DLA}
Adrian Sch\"{u}pbach, Andrew Baumann, Timothy Roscoe, and Simon Peter.
\newblock {A Declarative Language Approach to Device Configuration}.
\newblock In {\em Proceedings of the Sixteenth International Conference on
  Architectural Support for Programming Languages and Operating Systems},
  ASPLOS XVI, pages 119--132, New York, NY, USA, 2011. ACM.

\bibitem{Sewell:2011:Integrity}
Thomas Sewell, Simon Winwood, Peter Gammie, Toby Murray, June Andronick, and
  Gerwin Klein.
\newblock {seL4 Enforces Integrity}.
\newblock In Markovan Eekelen, Herman Geuvers, Julien Schmaltz, and Freek
  Wiedijk, editors, {\em Interactive Theorem Proving}, pages 325--340, Berlin,
  Heidelberg, 2011. Springer Berlin Heidelberg.

\bibitem{Vermij:2017:AIN}
Erik Vermij, Leandro Fiorin, Rik Jongerius, Christoph Hagleitner, Jan~Van
  Lunteren, and Koen Bertels.
\newblock An architecture for integrated near-data processors.
\newblock {\em ACM Trans. Archit. Code Optim.}, 14(3):30:1--30:25, September
  2017.

\bibitem{Volos:2018:GTE}
Stavros Volos, Kapil Vaswani, and Rodrigo Bruno.
\newblock Graviton: Trusted execution environments on gpus.
\newblock In {\em Proceedings of the 12th USENIX Conference on Operating
  Systems Design and Implementation}, OSDI’18, page 681–696, USA, 2018.
  USENIX Association.

\bibitem{lwn:2005:linuxsparsemem}
Andy Whitcroft.
\newblock {Sparsemem Memory Model}.
\newblock {\url{https://lwn.net/Articles/134804/}}, 8 2019.

\bibitem{Winwood:2009:MG}
Simon Winwood, Gerwin Klein, Thomas Sewell, June Andronick, David Cock, and
  Michael Norrish.
\newblock {Mind the Gap}.
\newblock In {\em Proceedings of the 22nd International Conference on Theorem
  Proving in Higher Order Logics}, TPHOLs '09, pages 500--515, Berlin,
  Heidelberg, 2009. Springer-Verlag.

\bibitem{Zhang:2014:TTP}
Dongping Zhang, Nuwan Jayasena, Alexander Lyashevsky, Joseph~L. Greathouse,
  Lifan Xu, and Michael Ignatowski.
\newblock Top-pim: Throughput-oriented programmable processing in memory.
\newblock In {\em Proceedings of the 23rd International Symposium on
  High-performance Parallel and Distributed Computing}, HPDC '14, pages 85--98,
  New York, NY, USA, 2014. ACM.

\bibitem{Zhu:2017:USD}
Zhiting Zhu, Sangman Kim, Yuri Rozhanski, Yige Hu, Emmett Witchel, and Mark
  Silberstein.
\newblock Understanding the security of discrete gpus.
\newblock In {\em Proceedings of the General Purpose GPUs}, GPGPU-10, pages
  1--11, New York, NY, USA, 2017. ACM.

\end{thebibliography}

\end{document}